\documentclass{aa} 

\usepackage{graphicx}
\usepackage{txfonts}
\usepackage{amsmath}
\usepackage{physics} 	
\usepackage{color} 		
\usepackage{verbatim} 	
\usepackage{threeparttable} 
\usepackage{epstopdf}
\usepackage{natbib}
\usepackage[colorlinks,linkcolor=blue,urlcolor=blue,citecolor=blue]{hyperref} 
\usepackage{subcaption} 

\usepackage{caption} 
\definecolor{Blue}{rgb}{0,0.08,0.9}
\definecolor{Red}{rgb}{0.85,0.08,0.05}
\definecolor{Green}{rgb}{0.35,0.45,0.25}
\definecolor{Orange}{rgb}{1.0,0.5,0.15}
\definecolor{Yellow}{rgb}{1.,1.,0.}
\definecolor{Brown}{rgb}{0.7,0.25,0.0}

\def\red{\color{Red}}

\def\olivierK#1{\noindent{\bf\red[$\diamondsuit$ #1]}}

\begin{document}  

   \title{Simulating \textit{JWST} deep extragalactic imaging surveys and physical parameter recovery}
   \author{O. B. Kauffmann\inst{1} \and O. Le Fèvre\inst{1} \and O. Ilbert\inst{1} \and J. Chevallard\inst{2} \and C. C. Williams\inst{3} \and E. Curtis-Lake\inst{4,5} \and L. Colina\inst{6,7} \and P. G. Pérez-González\inst{6} \and J. P. Pye\inst{8} \and K. I. Caputi\inst{9}
          }
   \institute{
   Aix Marseille Univ, CNRS, LAM, Laboratoire d'Astrophysique de Marseille, Marseille, France
   \and Sorbonne Universités, UPMC-CNRS, UMR7095, Institut d’Astrophysique de Paris, F-75014, Paris, France 
   \and Steward Observatory, University of Arizona, 933 North Cherry Avenue, Tucson, AZ 85721, USA 
   \and Cavendish Astrophysics, University of Cambridge,  Cambridge, CB3 0HE, UK
   \and Kavli Institute for Cosmology, University of Cambridge, Madingley Road, Cambridge CB3 0HA, UK
   \and Centro de Astrobiología, Departamento de Astrofísica, CSIC-INTA, Cra. de Ajalvir km.4, E-28850—Torrejón de Ardoz, Madrid, Spain 
   \and International Associate, Cosmic Dawn Center (DAWN) at the Niels Bohr Institute, University of Copenhagen and DTU-Space, Technical University of Denmark, Copenhagen, Denmark 
   \and University of Leicester, School of Physics \& Astronomy, Leicester, LE1 7RH, UK 
   \and Kapteyn Astronomical Institute, University of Groningen, P.O. Box 800, 9700 AV, Groningen, The Netherlands 
   \\\email{olivier.kauffmann@lam.fr}
          }   
   \date{Preprint online version: \today}

  \abstract
   {
We present a new prospective analysis of deep multi-band imaging with the \textit{James Webb} Space Telescope (\textit{JWST}).
In this work, we investigate the recovery of high-redshift $5<z<12$ galaxies through extensive image simulations of accepted \textit{JWST} programs such as CEERS in the EGS field and HUDF GTO.
We introduce complete samples of $\sim$300,000 galaxies with stellar masses $\log(M_*/M_\odot)>6$ and redshifts $0<z<15$, as well as galactic stars, into realistic mock NIRCam, MIRI and \textit{HST} images to properly describe the impact of source blending. 
We extract the photometry of the detected sources as in real images and estimate the physical properties of galaxies through spectral energy distribution fitting. 
We find that the photometric redshifts are primarily limited by the availability of blue-band and near-infrared medium-band imaging.
The stellar masses and star-formation rates are recovered within 0.25 and 0.3~dex respectively, for galaxies with accurate photometric redshifts.
Brown dwarfs contaminating the $z>5$ galaxy samples can be reduced to $<0.01$~arcmin$^{-2}$ with a limited impact on galaxy completeness.
We investigate multiple high-redshift galaxy selection techniques and find the best compromise between completeness and purity at $5<z<10$ using the full redshift posterior probability distributions.
In the EGS field, the galaxy completeness remains higher than $50\%$ for $m_\text{UV}<27.5$ sources at all redshifts, and the purity is maintained above 80 and 60\% at $z\leq7$ and 10 respectively.
The faint-end slope of the galaxy UV luminosity function is recovered with a precision of 0.1-0.25, and the cosmic star-formation rate density within 0.1~dex. We argue in favor of additional observing programs covering larger areas to better constrain the bright end.

    }

   \keywords{
   galaxies: high-redshift -- galaxies: photometry -- galaxies: distances and redshifts -- galaxies: fundamental parameters -- galaxies: evolution
               }

   \maketitle
%

\section{Introduction}

The detection of distant sources has been mainly driven by multi-wavelength photometry, through deep imaging over selected areas of the sky.
The \textit{Hubble} Space Telescope (\textit{HST}), with its Advanced Camera for Surveys (ACS) and Wide-Field Camera~3 (WFC3), enabled the discovery of many high-redshift galaxies with its deep optical and near-infrared imaging \citep[e.g.,][]{scoville_cosmic_2007,koekemoer_cosmos_2007,wilkins_ultraviolet_2011,schenker_uv_2013,mclure_new_2013,bouwens_uv_2015}, effectively covering the rest-frame UV region of these sources. 
Near-infrared observations are necessary to detect high-redshift galaxies because of the strong attenuation by the IGM blueward the Lyman limit, and as the Universe becomes more neutral, the flux blueward Lyman alpha also gets attenuated. 
The mid-infrared observations with the \textit{Spitzer} Space Telescope improved the characterization of galaxy physical properties 
that are required to constrain galaxy evolution from the epoch of reionization to the present day \citep[e.g.,][]{sanders_s-cosmos:_2007,caputi_spitzerbright_2015}. In particular, \textit{Spitzer} provided most of the constraints on the rest-frame optical at high redshift \citep{oesch_most_2014,oesch_dearth_2018}. 
The census of high-redshift sources is particularly important to estimate which sources contributed most of the ionizing photons necessary to support neutral hydrogen reionization. The latest accounts point to a high number of faint sources producing enough ionising photons \citep{bouwens_uv_2015,robertson_cosmic_2015} reconciling a late reionization supported by the latest CMB constraints on the Thomson scattering optical depth \citep{collaboration_planck_2018} and UV photons from galaxy counts \citep{madau_cosmic_2014}. Establishing a complete and unbiased census of galaxies and associated ionizing photons remains a priority to understand this important transition phase in the Universe, directly linked to the formation of the first galaxies.

Identifying high-redshift galaxies within, and following, the epoch of reionization $(5<z<12)$ is challenging because of their low number density which decreases with redshift.
The methods to select high-redshift candidates mostly rely on the identification of the dropout in continuum emission blueward Lyman alpha \citep{steidel_spectroscopic_1996}. 
Lyman break galaxies (LBG) can be identified through color-color selections, mainly using photometry in the rest-frame UV.
Alternatively, photometric redshifts obtained from spectral energy distribution (SED) fitting make use of all the photometric information \citep{finkelstein_evolution_2015}, spanning the optical to near-infrared, but do introduce model dependencies.
%
%
With the large number of low-redshift sources, that are several orders of magnitude more numerous, the high-redshift galaxy samples are subject to contamination because of the similar colors of these sources in the observed frame.
The main contaminants are low-redshift ($z\sim1-2$) dust obscured galaxies with very faint continuum in the visible bands \citep{tilvi_discovery_2013}. 
Brown dwarfs are other potential contaminants of the $z>5$ galaxy samples because of their similar near-infrared colors. 
%
The number of detected sources increases with telescope sensitivity, which naturally leads to an increasing probability of finding multiple objects along the line-of-sight, so that the impact of source blending becomes more important \citep{dawson_ellipticity_2016}.
In the case of source confusion, the background estimation becomes more challenging and individual sources harder to isolate. 
The background level by itself also affects source separation, so that extended sources with internal structures may be mistaken for multiple nearby objects.
In addition, the galaxy morphology is more complex at high redshifts \citep{ribeiro_size_2016}, therefore requiring adapted source detection techniques.

The \textit{James Webb} Space Telescope (\textit{JWST}\footnote{\url{http://www.stsci.edu/jwst}}, \citealt{gardner_science_2006}), to be launched in 2021, will revolutionize near- and mid-infrared astronomy. It will provide the first sub-arcsecond high-sensitivity space imaging ever at wavelengths above 3~microns and up to 25~microns, overcoming the current limitations of ground-based and space-based observatories. 
The on-board instruments include two imaging cameras, the Near-Infrared Camera (NIRCam\footnote{\url{https://jwst-docs.stsci.edu/near-infrared-camera}}, \citealt{rieke_overview_2005}) and the Mid-Infrared Instrument (MIRI\footnote{\url{https://jwst-docs.stsci.edu/mid-infrared-instrument}}, \citealt{rieke_mid-infrared_2015,wright_mid-infrared_2015}), 
together covering the wavelength range from 0.6 to 28~microns. These capabilities are perfectly suited to the discovery and the study of high-redshift galaxies during the epoch of reionization at $z>6$, in combination with the deep optical imaging from \textit{HST} and other ancillary data.

Predictions are required for preparation of the deep \textit{JWST} imaging programs. The observed number counts per field of view and their redshift distribution need to be quantified, as well as the source detectability and the completeness and purity of the selected samples, depending on the detection method. 
The most direct number count predictions require the integral of the luminosity function multiplied by the differential comoving volume over a given area and redshift interval. High-redshift luminosity functions may be estimated by either extrapolating some lower-redshift measurements or using semi-analytic modeling \citep{mason_galaxy_2015,furlanetto_minimalist_2017,cowley_predictions_2018,williams_jwst_2018,yung_semi-analytic_2019}. 
These methods quantify the expected number of detectable sources in a given field, not the number of sources that may be extracted and correctly characterized.
%
%
Alternatively, the recovery of the galaxy physical parameters may be simulated with mock galaxy photometry and SED-fitting procedures. \citet{bisigello_impact_2016} tested the derivation of galaxy photometric redshifts with \textit{JWST} broad-band imaging, considering multiple combinations of NIRCam, MIRI and ancillary optical bands. The galaxy physical parameter recovery was investigated using the same methodology \citep{bisigello_recovering_2017,bisigello_statistical_2019}.
Analogously, \citet{kemp_maximizing_2019} analyzed of the posterior constraints on the physical properties from SED-fitting with \textit{JWST} and \textit{HST} imaging.

\begin{figure*}
	\centering
	\includegraphics[width=0.65\hsize]{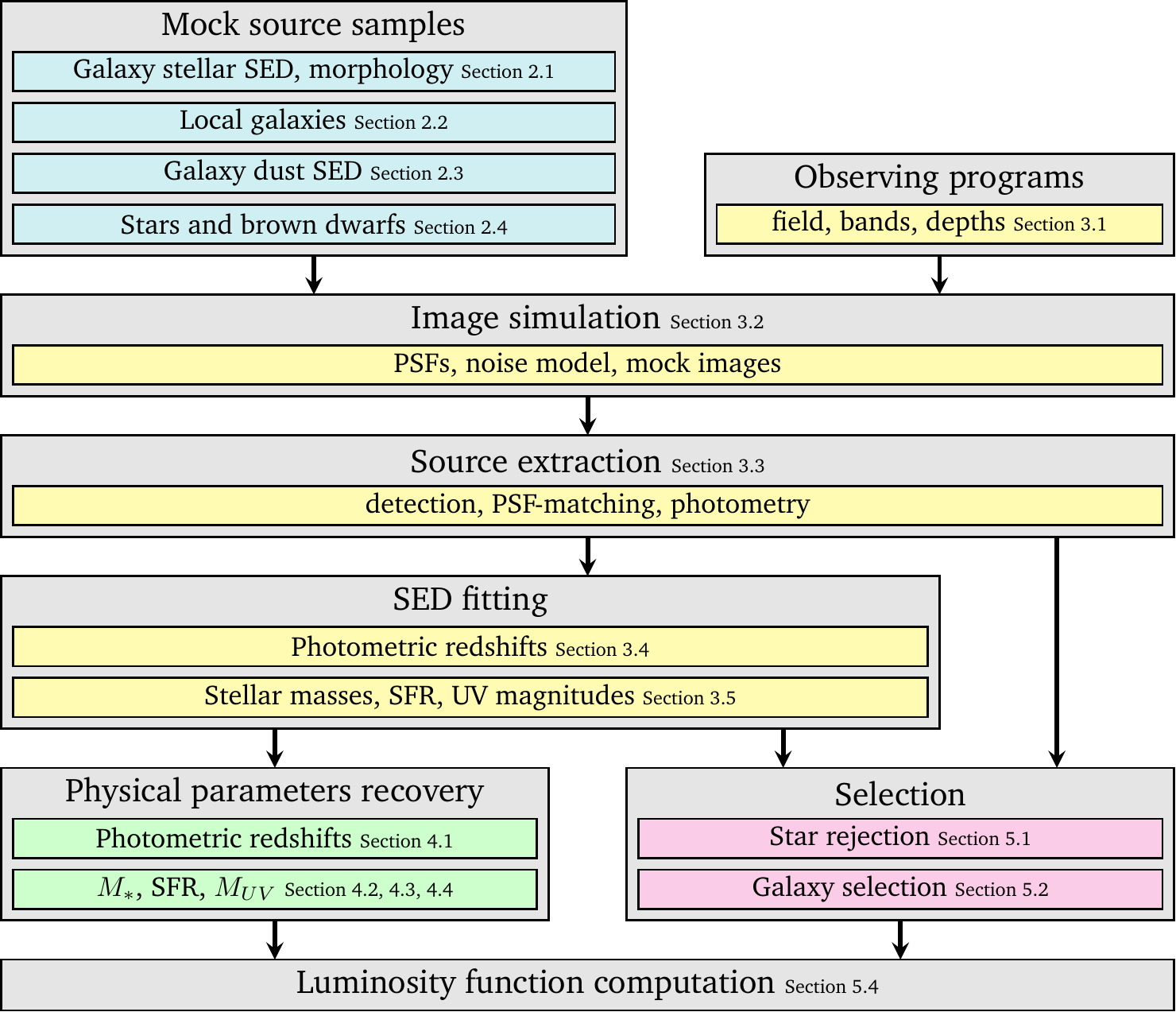}
	\caption{Diagram summarizing the procedures to make our predictions. The gray boxes indicate the essential steps, and colored boxes show the detail of the subsections. The colors code for the main sections.}
	\label{fig:diagram}
\end{figure*}

The aim of this paper is to investigate how to best identify high-redshift galaxies in the redshift range $5<z<12$ from \textit{JWST} deep-field imaging, to estimate their number counts, with associated completeness and purity, and how their physical parameters can be recovered, focusing on stellar mass ($M_*$) and star-formation rate (SFR). 
We concentrate on the identification and characterization of high-redshift sources from photometry, which will be required to identify sources for spectroscopic follow-up with \textit{JWST} (NIRSpec, \citealt{birkmann_jwst/nirspec_2016}). 
The simulation of deep fields necessitates the construction of realistic mock samples of sources, including all galaxies at all redshifts, as well as stars from the Galaxy. 
Any contamination estimate relies on the ability to produce simulations with sources that have realistic distributions of physical properties as a function of redshift, including fluxes and shapes projected on the image plane, as currently documented. 
In determining magnitudes, we need to include emission lines with strength corresponding to what is actually observed.
In this way the contamination of high-redshift galaxy samples by low-redshift interlopers and Galactic stars can be estimated. We neglect quasars and transient objects. 
Existing observations are not deep enough to use as a basis for predictions for \textit{JWST} and therefore some extrapolations are needed. 
%
To take geometrical effects into account, we generate mock images from the current knowledge of the instruments, then extract and identify  sources. This allows us to more realistically characterize the statistical properties of the galaxy population, and especially source blending, thanks to the complete source sample. 
Figure~\ref{fig:diagram} summarizes our methodology to make our forecasts.


This paper is organized as follows. In Sect.~\ref{section2} we present the mock source samples, including galaxies and stellar objects. Section~\ref{section3} describes our methodology to simulate images, extract sources and measure photometry and physical parameters. The results of the physical parameter recovery are detailed in Sect.~\ref{section4}. Section~\ref{section5} describes our source selection investigations, including the rejection of the stellar contaminants, high-redshift galaxy selection and luminosity function computation. We summarize and conclude in Sect.~\ref{section6}.
Magnitudes are given the AB system \citep{oke_absolute_1974}, and we adopt the standard $\Lambda$CDM cosmology with $\Omega_\text{m}=0.3$, $\Omega_\Lambda=0.7$ and $H_0=70$~km s$^{-1}$ Mpc$^{-1}$.

\section{Mock source samples}
\label{section2}

\subsection{Mock galaxy sample}
\label{subsection_galaxy_sample}

We build our galaxy sample from the JADES extraGalactic Ultradeep Artificial Realizations v1.2 (JAGUAR\footnote{\url{http://fenrir.as.arizona.edu/jwstmock/}}, \citealt{williams_jwst_2018}) developed for the \textit{JWST} Advanced Deep Extragalactic Survey (JADES). 
This phenomenological model of galaxy evolution generates mock galaxy catalogs with physical and morphological parameters, reproducing observed statistical functions. 
Publicly available realizations consist of complete samples of star-forming and quiescent galaxies with stellar mass $6<\log(M_*/M_\odot)<11.5$ and redshift $0.2<z<15$ on areas of $11\times11$~arcmin$^2$, each containing $\sim 3\times10^5$ sources.

\begin{figure}
	\centering
	\includegraphics[width=\hsize]{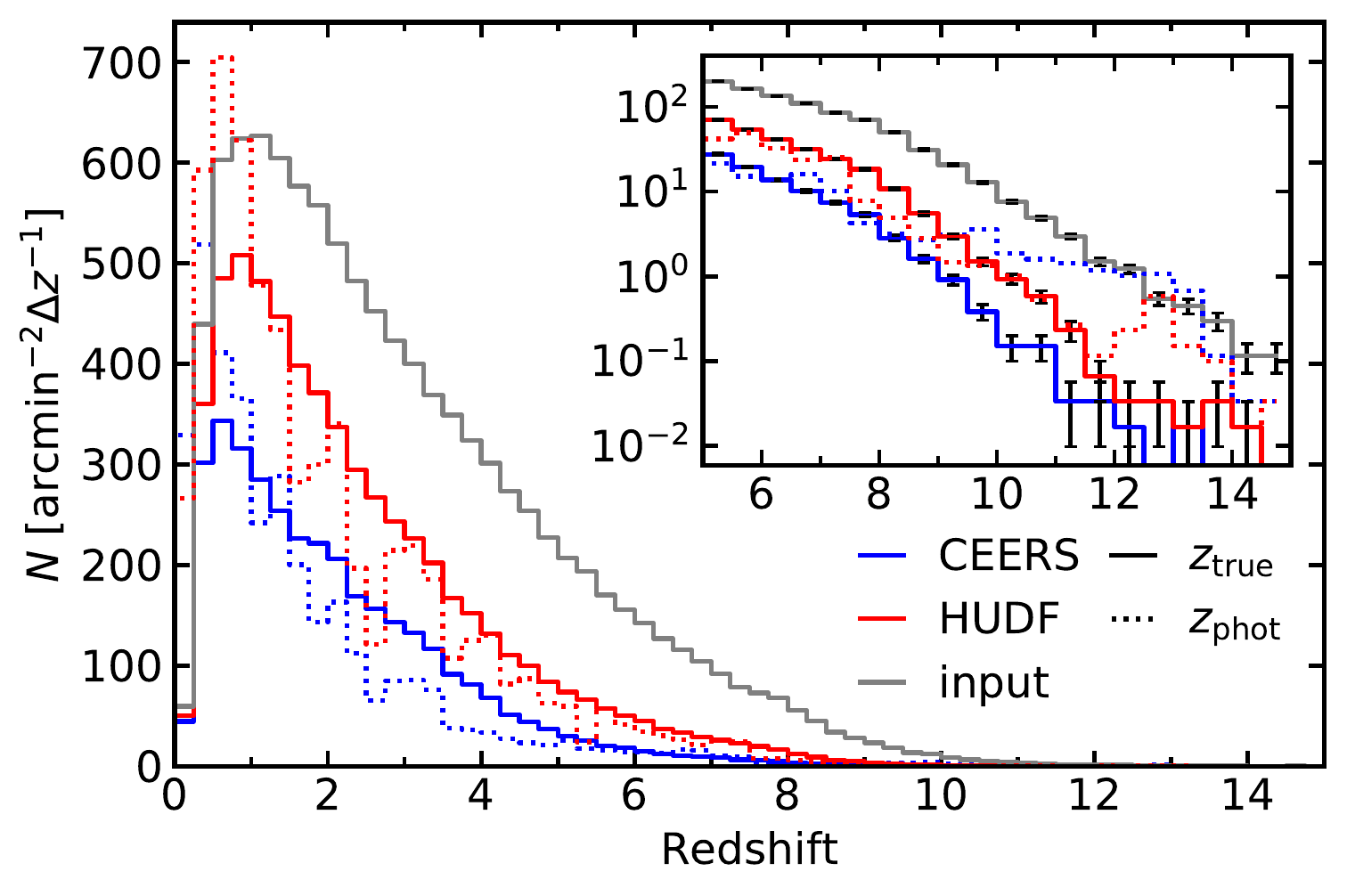}
	\caption{Galaxy surface number density
	versus redshift. 
	The gray line includes all the mock galaxies with stellar masses $\log(M_*/M_\odot)>6$. The colored lines illustrate the redshift distribution of the detected sources in our CEERS and HUDF simulations. The solid and dashed lines represent input and photometric redshift distributions, respectively. The inset provides a zoom-in at high redshift, with Poisson error bars. 
	}
	\label{fig_Nz_input}
\end{figure}

Stellar masses and redshifts are sampled from a continuous stellar mass function (SMF) model, constructed from the empirical SMF constraints of \citet{tomczak_galaxy_2014} at $z<4$ and the luminosity functions (LF) of \citet{bouwens_uv_2015} and \citet{oesch_dearth_2018} at $z<10$.
We note that these observations support a rapid evolution of the UV LF at $z>8$ inducing a strong decrease in galaxy number counts, which is still debated in the literature \citep[e.g.,][]{mcleod_new_2015,mcleod_z_2016}. The SMF model separately describes star-forming and quiescent galaxies and is extrapolated beyond $z=10$. 
Physical parameters (e.g., UV magnitude $M_\text{UV}$, UV spectral slope $\beta$) are sampled from observed relationships and their scatter between $M_\text{UV}$ and $M_*$, and $M_\text{UV}$ and $\beta$. 
A spectral energy distribution (SED) is assigned to each set of physical parameters using BEAGLE \citep*{chevallard_modelling_2016}. 
\citet{williams_jwst_2018} describe a galaxy star-formation history (SFH) with a delayed exponential function and model stellar emission with the latest version of the \citet*[hereafter BC03]{bruzual_stellar_2003} population synthesis code. They consider the (line+continuum) emission from gas photo-ionized by young, massive stars using the models of \citet{gutkin_modelling_2016}. 
Dust attenuation is described by the two-component model of \citet*{charlot_simple_2000} and parametrized in terms of the $V-$band attenuation optical depth $\hat{\tau}_V$ and the fraction of attenuation arising from the diffuse ISM (set to $\mu=0.4$), while IGM absorption follows the prescriptions of \citet{inoue_updated_2014}.
No galaxies composed of metal-free population III stars are considered because of the lack of knowledge about these objects. No AGN models are considered either.
Figure~\ref{fig_Nz_input} shows the redshift distribution of the mock galaxy catalog.

Galaxy morphology is parametrized by one Sérsic function \citep{sersic_influence_1963} assumed to be wavelength-independent.
Effective radii are sampled from a continuously evolving model with stellar mass and redshift, based on the observed size-mass relations in CANDELS and the 3D-\textit{HST} survey by \citet{van_der_wel_3d-hst+candels:_2014}.
This model separately treats star-forming and quiescent galaxies, and extrapolates the observed trends down to $\log(M_*/M_\odot)=6$. 
Axis ratio and Sérsic indices are sampled from the redshift-dependent distributions of \citet{van_der_wel_structural_2012}. 
The description of galaxy surface brightness profiles relies on observed UV magnitudes and apparent shape measurements, so that only the strong lensing shape distortions are neglected here. Magnification is naturally included, although we do not expect magnification bias to be important in \textit{JWST} pencil-beam surveys containing few $M_\text{UV}<-22$ galaxies \citep{mason_correcting_2015}.
%
The underlying assumptions on the morphology of galaxies, especially at $z>2$, may have important consequences on whether a source can be recovered, with two main limitations. Size measurements which assume symmetry find that sizes decrease with redshift (\citealt{shibuya_morphologies_2015}, although cf. \citealt{curtis-lake_non-parametric_2016}), whereas one finds that sizes remain large and constant with redshift when more adapted isophotal limits are used \citep{ribeiro_size_2016}. This arises from galaxies becoming more complex, multi-component as redshift increases, therefore spreading the total flux over a large area with surface brightness becoming lower.
In addition, a clumpy galaxy may be resolved at certain wavelengths but appear mono-component in others because of the change of intrinsic structure and/or the varying angular resolution. This could be an increasing problem with source identification and multi-wavelength photometry. 
While this work is based on symmetric profiles, we will investigate how multi-component galaxies can be detected in a future paper.


Galaxy coordinates are sampled from a uniform distribution over the surveyed area. We therefore neglect galaxy alignment from clustering or lensing.
The position of galaxy pairs with large line-of-sight separations are independent, meaning that the blending of high-redshift galaxies with low-redshift sources should remain unchanged with and without clustering, at the first order. Multiple over-dense regions may happen to be on the same line of sight, however we neglect these cases.
Moreover, we only simulate non-confused, high-resolution imaging, so that we expect our predictions to be negligibly impacted by clustering. 



Because of the rarity of the high-redshift sources, large areas of the sky need to be simulated to make predictions with sufficient statistical significance. 
We replicate three times the initial galaxy catalog covering $11\times11$~arcmin$^2$ to increase the simulated area and sample size. For each replication, stellar masses are sampled from a centered log-normal distribution with $\sigma=0.1$~dex. This ensures that the difference in the resulting SMF remains below $\sim$10\% at $M_*<10^{11}$~$M_\odot$. Fluxes and SFRs are modified accordingly. Coordinates and galaxy position angles are randomized, and the other parameters kept unchanged.

\subsection{Local galaxies}

Low-redshift galaxies are one of the main sources of contamination for the high-redshift galaxy samples, notably because of the degeneracy between the Lyman and the Balmer breaks \citep[e.g.,][]{le_fevre_vimos_2015}.
Pencil-beam surveys contain very few local galaxies, however the apparent size of these objects are the largest on the sky, so that it is important to take them into account when simulating realistic blending.

The JAGUAR galaxy catalog does not include $0<z<0.2$ galaxies, because of the lack of low-redshift volume considered in building the stellar mass function in \citet{tomczak_galaxy_2014}. We sample redshifts and stellar masses at $M_*>10^6$~$M_\odot$ from the SMF continuous model of \citet{wright_gama/g10-cosmos/3d-hst:_2018} to fill this redshift interval. In that paper, the authors made use of the GAMA (60~deg$^2$), G10-COSMOS (1~deg$^2$) and 3D-\textit{HST} (0.274~deg$^2$) data set gathered by \citet{driver_gama/g10-cosmos/3d-hst:_2018} to efficiently constrain both the bright and the faint ends of the SMF. For comparison, the area of the data used in \citealt{tomczak_galaxy_2014} is 316~arcmin$^2$.
We sample about 460 (50) galaxies with $\log(M_*/M_\odot)>6$ (8) over $11\times11$~arcmin$^2$. We assign the spectrum from the JAGUAR $0.2<z<0.4$ galaxy with the closest stellar mass to each sampled parameter set. The maximum stellar age in these galaxies is therefore underestimated. By construction, about half of these galaxies are quiescent. The morphological parameters are sampled from the same distributions as in JAGUAR.



\subsection{Galaxy infrared spectra}

Dust emission can make a significant contribution to the near- and mid-infrared galaxy spectrum. In addition, mid-infrared photometry may considerably help to identify low-redshift contaminants to high-redshift samples using photometric redshift estimation \citep[e.g.,][]{ilbert_cosmos_2009}. 
Because the galaxy spectra in JAGUAR include stellar and nebular emission, we include the additional dust emission for a more accurate modeling of the galaxy mid-infrared spectra.
We neglect dust emission for low-mass\footnote{with $\log(M_*/M_\odot)<8.7+0.4z$} quiescent galaxies because \citet{williams_jwst_2018} neglected dust attenuation for these objects.

We take the library of dust spectral energy distributions of \citet{schreiber_dust_2018} constructed from the dust models of \citet{galliano_non-standard_2011}. These templates separately describe the dust grain continuum emission and the polycyclic aromatic hydrocarbon (PAH) emission. The contribution of an AGN torus to the dust emission is neglected. The dust temperature ($T_\text{dust}$) determines the shape of both components, the mid-to-total infrared color ($\text{IR8}=L_\text{IR}/L_{8\mu\text{m}}$) sets their relative contributions and the infrared luminosity ($L_\text{IR}$) scales the sum. 

\begin{figure}
	\centering
	\includegraphics[width=\hsize]{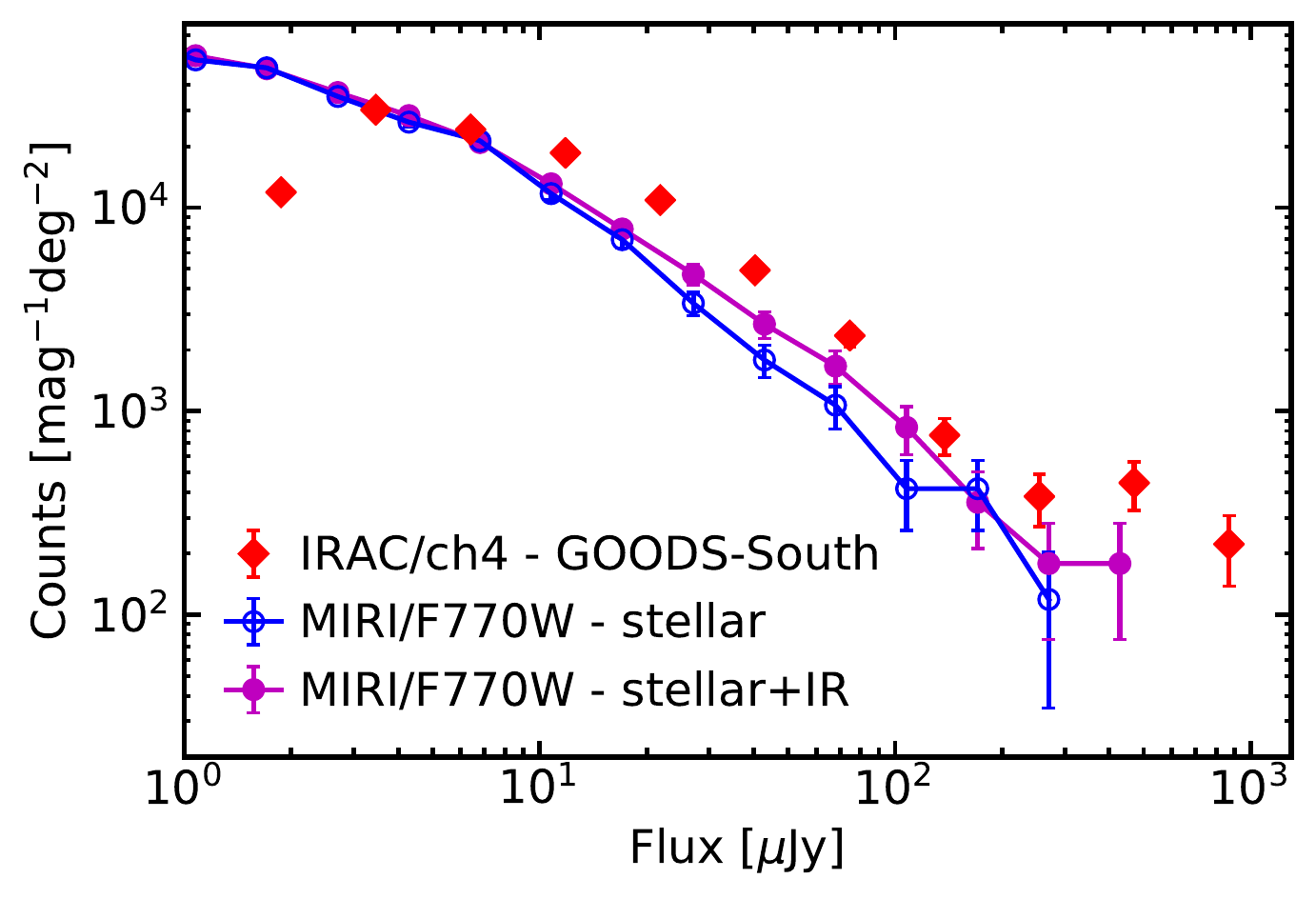}
	\caption{Differential galaxy number counts in the MIRI/F770W filter with and without the dust emission, compared to the \textit{Spitzer} IRAC/8~$\mu$m number counts measured in the GOODS-South field \citep{schreiber_egg:_2017}.
	}
	\label{fig_IRcounts_8}
\end{figure}

We attribute $T_\text{dust}$ and IR8 to all the mock galaxies following the empirical laws evolving with redshift from \citet{schreiber_dust_2018}, including the intrinsic scatter. These relations were calibrated from the stacked \textit{Spitzer} and \textit{Herschel} photometry \citep{schreiber_herschel_2015}.
We estimate the infrared luminosities from the $V-$band attenuation optical depth $\hat{\tau}_V$, assuming that the absorbed flux is entirely re-emitted by the dust (energy balance). We neglect the birth clouds component of the \citet{charlot_simple_2000} attenuation curve, since the JAGUAR catalog only provides the summed emission from young and old stars. This may lead to underestimated dust emission, as well as the limitation to the diffuse ISM. Figure~\ref{fig_IRcounts_8} 
indicates a better agreement between simulated and empirical counts in the MIRI/F770W filter.

\subsection{Mock star sample}
\label{Mock star sample}

In this section, we present our formalism to generate mock stars from the Milky Way in the field of view.
The strategy to create the mock star catalog is the following: (1) estimate the number density per spectral type, (2) sample heliocentric distances and physical properties
, then (3) assign the spectrum with the closest properties.

We make use of the Besançon Model of the Galaxy\footnote{\url{http://model2016.obs-besancon.fr/}} 
\citep{robin_synthetic_2003,robin_stellar_2012,robin_constraining_2014} to generate mock stars of spectral type FGKM. This model of stellar population synthesis provides star samples with intrinsic parameters (mass, age, metallicity, effective temperature $T_\text{eff}$, surface gravity $\log g$). 
OBA stars are not sampled because of their rarity in pencil-beam surveys.
We follow the galaxy model of \citet{caballero_contamination_2008} to determine the mean number of LTY stars per unit area. The galaxy density profile is modeled by an exponential thin-disc with the parameters from \citet{chen_stellar_2001}, reliable at high galactic latitudes $b$. The surface density of objects at the central galactic coordinates $(l,b)$ results from the integration of the density profile over heliocentric distance, scaled to the local number density. 
We take the predicted local number densities of \citet{burgasser_binaries_2007} for L0 to T8 stars (see \citealt{caballero_contamination_2008}). Because of the small number of Y star observations, their number density is poorly constrained so we linearly extrapolate the local number densities of hotter stars to the cooler subtypes T9 and Y0-Y2. 
Star coordinates are sampled from a uniform distribution. 

\begin{figure}
	\centering
	\includegraphics[width=\hsize]{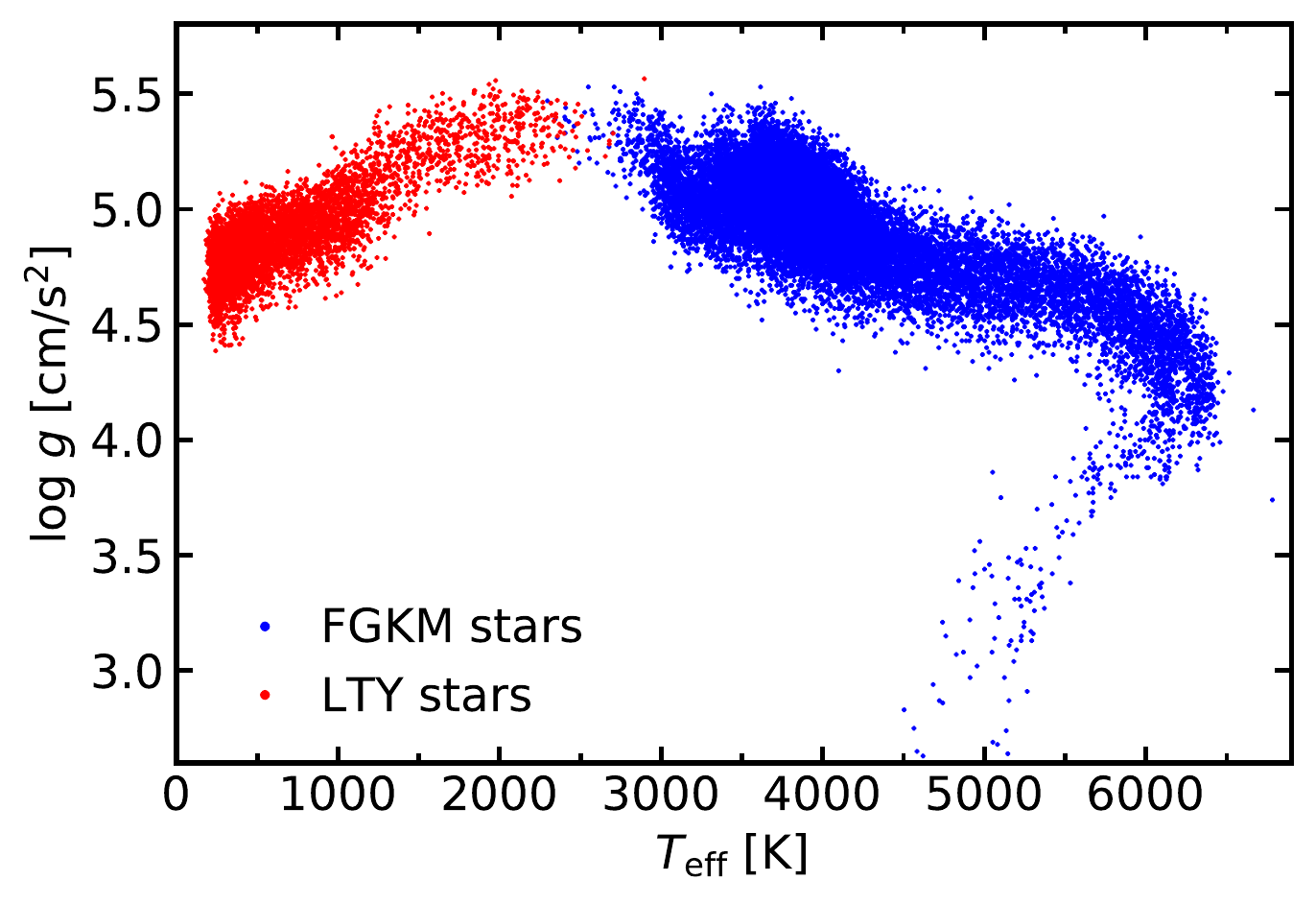}
	\caption{Distribution of effective temperature and surface gravity for the mock FGKM stars from the Besançon model in blue, and our mock LTY stars in red.
	}
	\label{fig_star_prop}
\end{figure}

An effective temperature $T_\text{eff}$ is assigned to each stellar subtype following the brown dwarf compilation\footnote{\url{http://www.pas.rochester.edu/~emamajek/}} from \citet*{pecaut_intrinsic_2013}. Surface gravity $\log g$ is computed from the stellar masses and radii, the latter being taken from the same compilation but imposing the lower bound of 0.1~$R_\odot$ (1 Jupiter radius). Stellar masses come from a linear model with 0.1~$M_\odot$ for type L0 and 0.02~$M_\odot$ for Y2. We include a bivariate Gaussian scatter to the duplet ($T_\text{eff}$, $\log g$) with 10\% relative dispersion. 
Figure~\ref{fig_star_prop} shows the sampled parameters for LTY stars. 
We sample the sedimentation efficiencies $f_\text{sed}$ from Gaussian random distributions with mean $\mu=2$ and scatter $\sigma=1$ for L types, and $\mu=4.5$ and $\sigma=0.5$ for T and Y types \citep{morley_neglected_2012}. This parameter describes the optical thickness of the metal clouds in the brown dwarf atmosphere.

We consider the modeled stellar spectra from \citealt{baraffe_new_2015} (BT-Settl, CIFIST2011\_2015, $1200<T_\text{eff}<7000$~K), \citealt{morley_neglected_2012} ($500<T_\text{eff}<1200$~K) and \citealt{morley_water_2014} ($200<T_\text{eff}<500$~K). These are physically-motivated high-resolution spectra from optical to mid-infrared, including absorption by water, methane, ammonia and metal clouds. 
We extrapolate the templates blueward 6000~\AA{} with a blackbody spectrum at the corresponding effective temperature if necessary. Cold brown dwarf spectra differ from blackbodies by several orders of magnitudes below 1~$\mu$m \citep{morley_water_2014}, hence we scale the blackbody spectrum to the bluest template point.
We assign to each parameter set ($T_\text{eff}$, $\log g$, $f_\text{sed}$) the template with the closest parameters. We check that our modeling can reproduce the optical and near-IR magnitudes from the Besançon model output for F to M stars. Emitted spectra are finally scaled according to the stellar radii and heliocentric distances. 


\section{Methodology}
\label{section3}

\subsection{Programs}

In this paper, we consider two accepted \textit{JWST} observing programs in the Extended Groth Strip (EGS) and the \textit{Hubble} Ultra-Deep Field (HUDF). The existing \textit{HST} imaging data in the optical and near-infrared are utilized in both fields. We exclusively simulate high-resolution space-based images and neglect ancillary ground-based data.

\begin{table*}[h]
\small
\centering
\renewcommand{\arraystretch}{1.1}
\begin{threeparttable}
\caption{Summary of the \textit{JWST} imaging data in CEERS and the HUDF -- Limiting magnitudes\tnote{a}}
\begin{tabular}{lcccccccccccc}
\hline\hline
name & area\tnote{b} & \multicolumn{9}{c}{NIRCam} & \multicolumn{2}{c}{MIRI} \\
 & [arcmin$^2$] & F090W & F115W & F150W & F200W & F277W & F335M & F356W & F410M & F444W & F560W & F770W \\\hline
CEERS\_1 & 96.8 & - & 28.7 & 28.9 & 29.1 & 29.2 & - & 29.2 & - & 28.7 & - & - \\
CEERS\_2 & 4.6 & - & 28.7 & 28.9 & 29.1 & 29.2 & - & 29.2 & - & 28.7 & 25.9 & 25.9 \\
HUDF\_1 & 4.7 & 29.9 & 30.3 & 30.3 & 30.3 & 30.6 & 29.9 & 30.5 & 29.9 & 30.1 & - & - \\
HUDF\_2 & 2.3 & 29.9 & 30.3 & 30.3 & 30.3 & 30.6 & 29.9 & 30.5 & 29.9 & 30.1 & 28.1 & - \\
\hline
\end{tabular}
\begin{tablenotes}
\item [a] The magnitudes are 5$\sigma$ point-source limits measured in 0.2" and 0.6" diameter apertures for NIRCam and MIRI, respectively.
\item [b] The configurations without MIRI include the area of the configurations with MIRI.
\end{tablenotes}
\label{tab:JWST-PSDL}
\end{threeparttable}
\end{table*}

\begin{table*}[h]
\small
\centering
\renewcommand{\arraystretch}{1.1}
\begin{threeparttable}
\caption{Summary of the \textit{HST} imaging data -- Limiting magnitudes\tnote{a}}
\begin{tabular}{lcccccccccc}
\hline\hline
field & area\tnote{b} & \multicolumn{5}{c}{ACS} & \multicolumn{4}{c}{WFC3} \\
 & [arcmin$^2$] & F435W & F606W & F775W & F814W & F850LP & F105W & F125W & F140W & F160W \\\hline
EGS & 205 & - & 28.8 & - & 28.2 & - & - & 27.6 & 26.8\tnote{c} & 27.6 \\
XDF & 4.7 & 29.8 & 30.3 & 30.3 & 29.1 & 29.4 & 30.1 & 29.8 & 29.8 & 29.8 \\
\hline
\end{tabular}
\begin{tablenotes}
\item [a] The magnitudes are 5$\sigma$ limits measured in empty circular apertures of diameter 2$\times$ the PSF FWHM.
\item [b] This corresponds to the WFC3 surveyed area.
\item [c] The WFC3/F140W band in the EGS field is not used in this paper (see Sect.~\ref{section:CEERS}).
\end{tablenotes}
\label{tab:HST-PSDL}
\end{threeparttable}
\end{table*}

\subsubsection{Cosmic Evolution Early Release Science survey}
\label{section:CEERS}

\begin{figure}[t]
	\centering
	\includegraphics[width=\hsize]{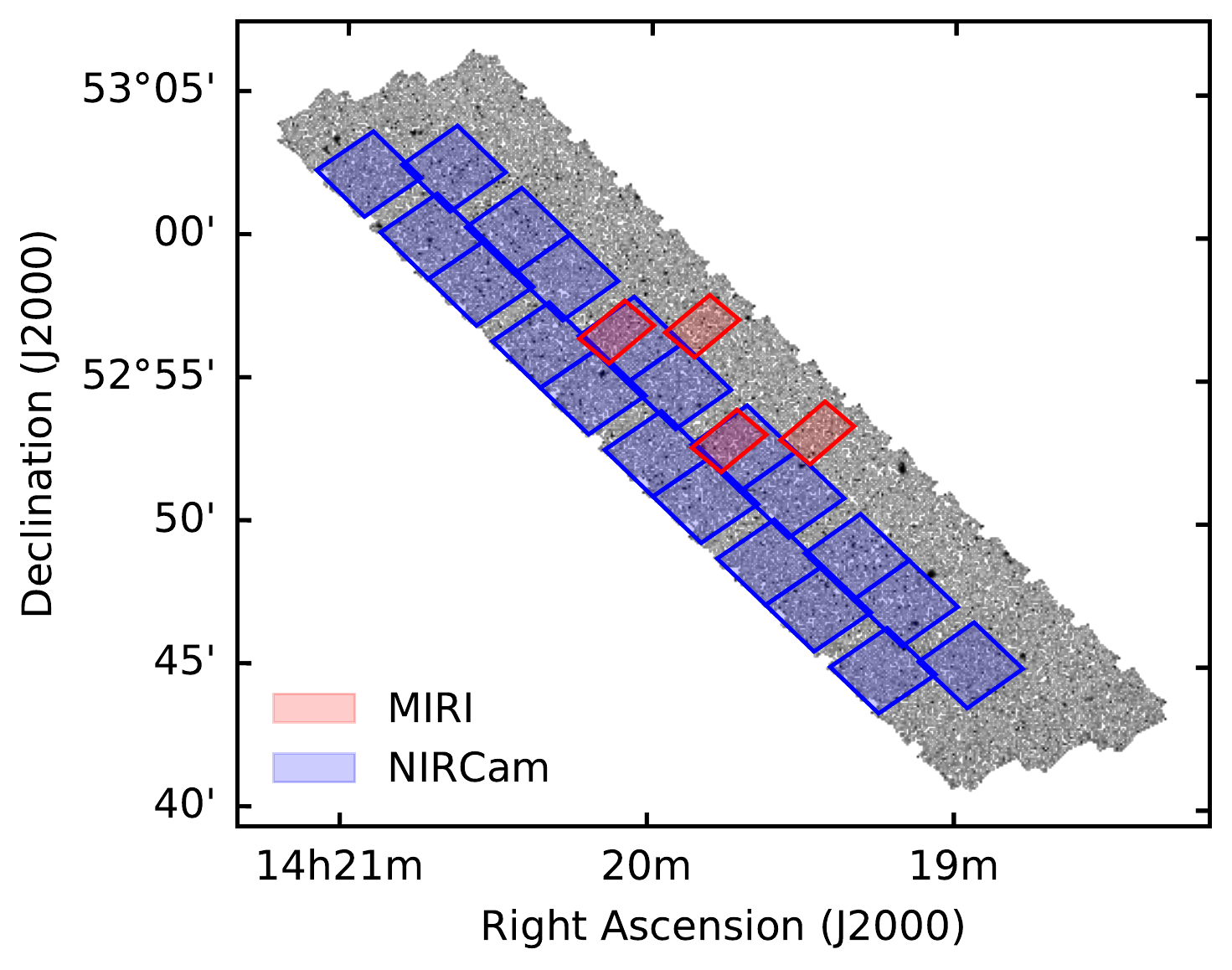}
	\caption{CEERS layout in the EGS field. The 10 NIRcam imaging pointings are shown in blue and the 4 MIRI parallels in red. The ancillary \textit{HST}/WFC3 $H_{160}$-band coverage is in gray. The pointings are all approximate until the final schedule. The parallel NIRSpec observations are not represented for clarity.}
	\label{fig_layout_EGS}
\end{figure}

The Cosmic Evolution Early Release Science (CEERS\footnote{\url{https://jwst.stsci.edu/observing-programs/approved-ers-programs}}; P.I.: S. L. Finkelstein) survey is one of the \textit{JWST} Early Release Science (ERS) programs. CEERS includes multiple imaging (NIRCam, MIRI) and spectroscopic observations over 100~arcmin$^2$ in the EGS \textit{HST} legacy field. As shown in Fig.~\ref{fig_layout_EGS}, the mosaic pattern consists of ten adjacent and non-overlapping NIRCam imaging pointings (each NIRCam pointing includes two parallel and separated fields), covering the $1<\lambda<5$~$\mu$m wavelength range, with four MIRI imaging parallels giving two NIRCam-MIRI overlaps. The estimated $5\sigma$ depths are $\sim29$~mag for NIRCam and $\sim26$~mag for MIRI (with 32~hours of science integration time). 
For simplicity, we treat the two distinct observing strategies listed in Table~\ref{tab:JWST-PSDL}. These are the shallowest NIRCam-only and NIRCam-MIRI configurations of the survey, though all the pointings have similar filter choices and exposure times.

The EGS field is supported by the \textit{HST}/CANDELS multi-wavelength data \citep{stefanon_candels_2017}. 
We consider the high-resolution \textit{HST} imaging in the ACS/F606W, F814W and WFC3/F125W, F160W bands \citep{grogin_candels:_2011,koekemoer_candels:_2011}, as indicated in Table~\ref{tab:HST-PSDL}. These images reach the $5\sigma$ depths of 28.8, 28.2, 27.6 and 27.6~mag respectively, measured in empty 0.24", 0.24", 0.38" and 0.4" diameter apertures. We do not use the WFC3/F140W imaging from 3D-HST \citep{brammer_3d-hst:_2012,momcheva_3d-hst_2016} because of its non-uniform layout.
In the future, the Ultraviolet Imaging of the CANDELS Fields (UVCANDELS; P.I.: H. Teplitz) will provide deep WFC3/F275W and ACS/F435W imaging in the EGS field, covering most of the WFC3 footprint and reaching about 27 and 28~mag depths, respectively. These data are not simulated either.

\subsubsection{\textit{Hubble} Ultra-Deep Field}

\begin{figure}[t]
	\centering
	\includegraphics[width=\hsize]{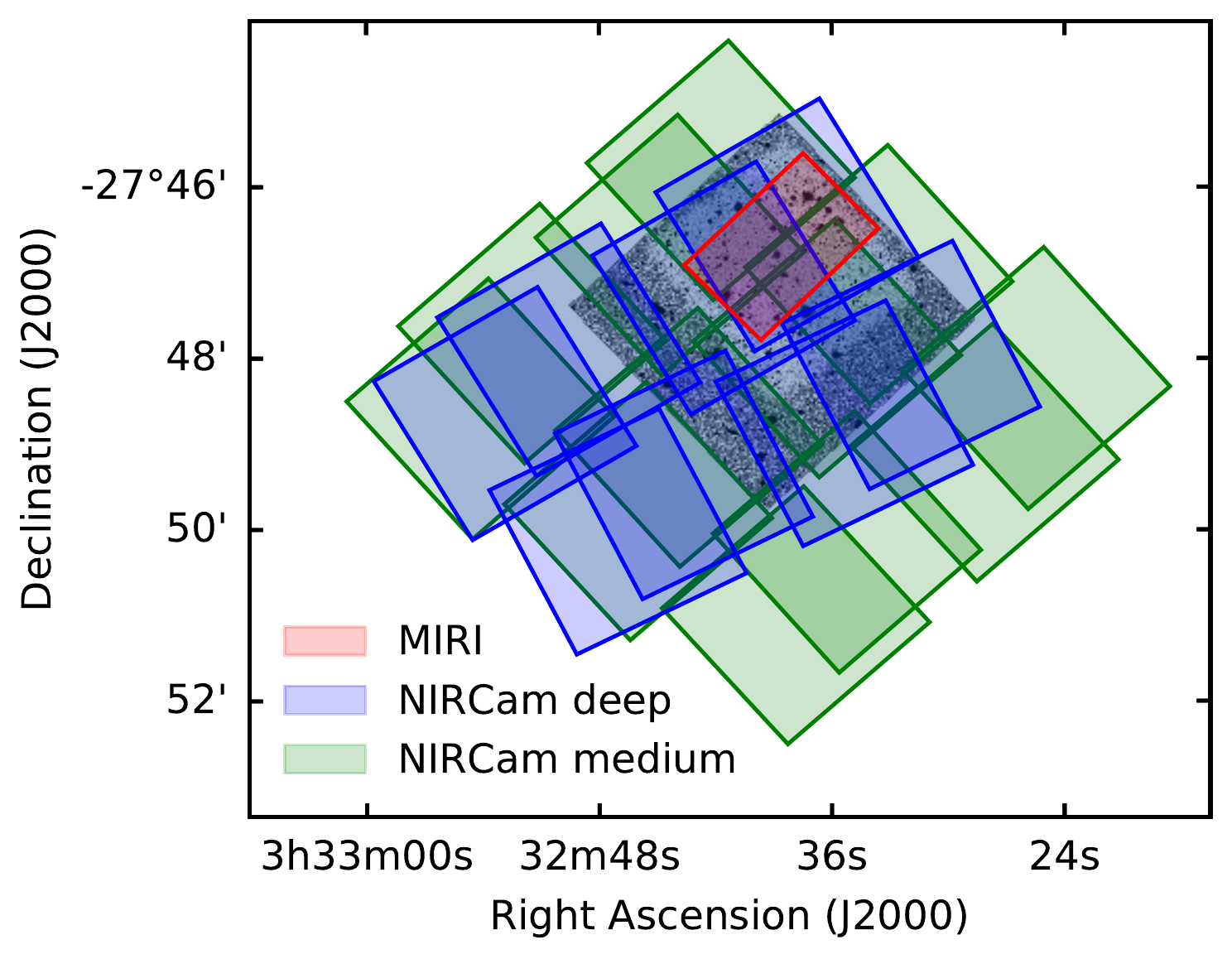}
	\caption{NIRCam GTO and MIRI GTO layouts in the HUDF. The 4 NIRCam deep pointings are shown in blue, the 6 NIRCam medium pointings in green and the MIRI pointing in red. The ancillary WFC3/IR $H_{160}$-band coverage in the XDF field is in gray, and the deepest WFC3 region in light gray. The whole XDF field is covered with the deepest ACS data. The dither patterns and the parallel observations are not represented for clarity.}
	\label{fig_layout_HUDF}
\end{figure}

Two programs of the Guaranteed Time Observations (GTO) teams are designed to observe the CANDELS GOODS-South field, both of them including deep imaging of the eXtreme Deep Field (XDF). The NIRCam-NIRSpec Galaxy Assembly Survey (P.I.: D.J. Eisenstein)
in the GOODS-South and GOODS-North fields includes deep NIRCam pre-imaging of the HUDF for spectroscopic follow-ups, separated into "Deep" and "Medium" pointings as shown in Fig.~\ref{fig_layout_HUDF}. 
This program covers $0.8<\lambda<5$~$\mu$m with broad-band imaging and two additional medium bands at 3.35 and 4.10~$\mu$m. In the GOODS-South field, the "Deep" ("Medium") survey covers 26 (40)~arcmin$^2$ with 174 (42) hours of science time integration, and consists of four (six) NIRCam pointings.
In both Deep and Medium pointings (separately), about one third of the area includes overlapping pointings.
With NIRCam alone, \citet{williams_jwst_2018} predicted several thousands of detected galaxies at $z>6$ and tens at $z>10$ at $<30$~mag ($5\sigma$) within the total $\sim200$~arcmin$^2$ survey in the GOODS-South and GOODS-North fields.
The MIRI HUDF Deep Imaging Survey (P.I.: H.U. Norgaard-Nielsen) 
consists of MIRI imaging in the F560W filter across the 2.3~arcmin$^2$ of the MIRI field of view. This survey will reach depths of $28.3$~mag (4$\sigma$) with 49 hours of science integration time for a total of 60 hours. Its layout will be entirely covered by NIRCam imaging.

\begin{figure}[t]
	\centering
	\includegraphics[width=\hsize]{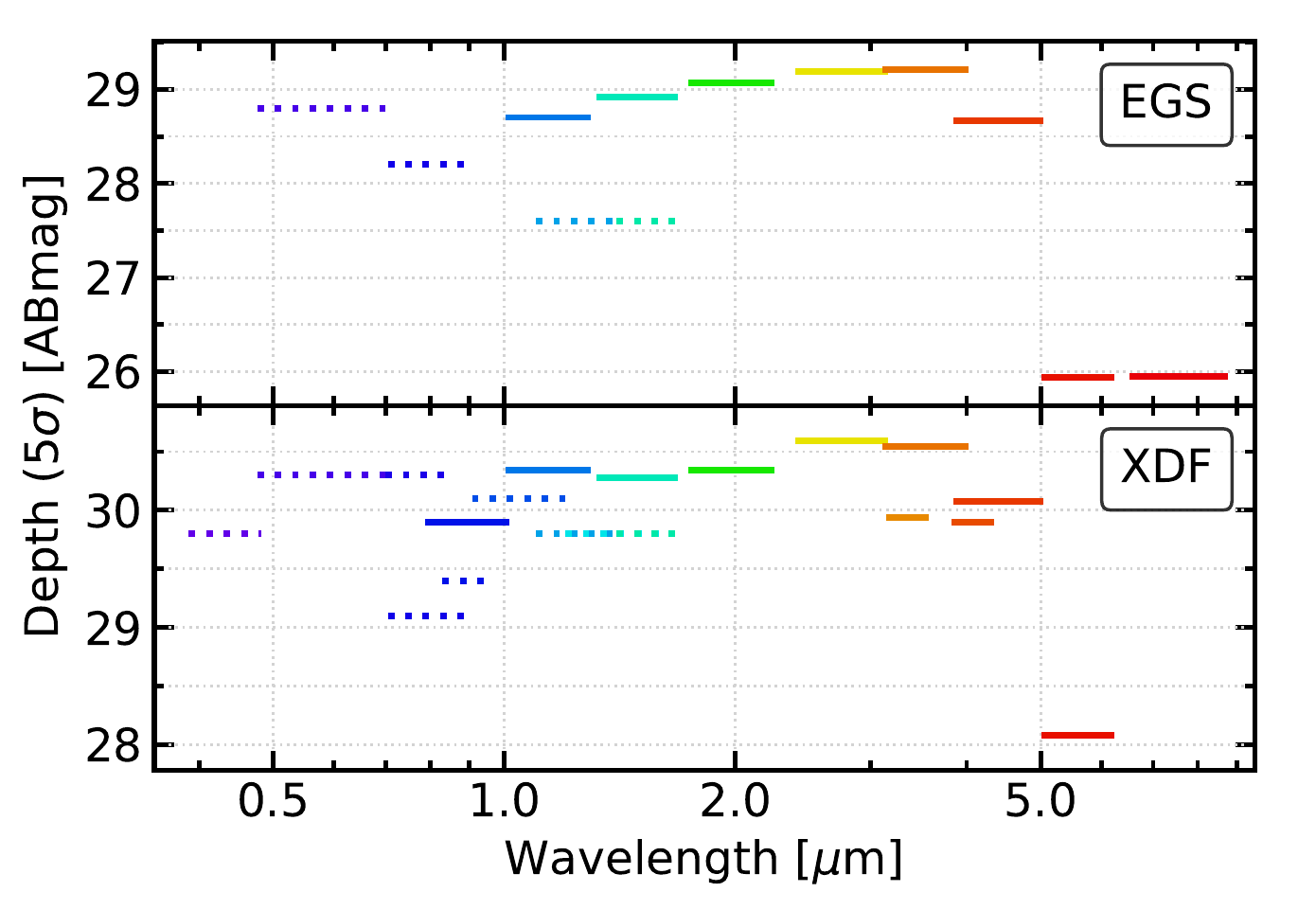}
	\caption{Limiting magnitudes at 5$\sigma$ in the simulated data sets in the EGS field (top) and the XDF (bottom). The list of bands and depths are listed in Tables~\ref{tab:JWST-PSDL} and \ref{tab:HST-PSDL}. The solid lines are \textit{JWST} bands and the dotted lines are \textit{HST} bands. The length of each segment is the FWHM of the filter transmission curve.}
	\label{fig_depth_comparison}
\end{figure}

This field benefits from existing \textit{HST}/ACS and WFC3/IR imaging, especially in the XDF with the deepest \textit{HST} imaging ever achieved \citep{illingworth_thehstextreme_2013,guo_candels_2013}. These images cover the optical and near-IR domain $0.38<\lambda<1.68$~$\mu$m with 9 filters, across 10.8 (4.7)~arcmin$^2$ for the deepest optical (infrared) data. In most filters, the typical $5\sigma$ depth reaches 30~mag in the deepest region, measured in empty 0.35" diameter apertures. 
We consider two configurations in the deepest WFC3/IR region as described in Table~\ref{tab:JWST-PSDL}, combining the NIRCam Deep and Medium pointings without the respective overlaps, and either with or without MIRI. The bands and depths are listed in Tables~\ref{tab:JWST-PSDL} and \ref{tab:HST-PSDL}, and represented in Fig.~\ref{fig_depth_comparison}.

\subsection{Mock image simulation}
\label{section3.2}

\begin{figure*}
	\centering
	\includegraphics[width=0.5\hsize]{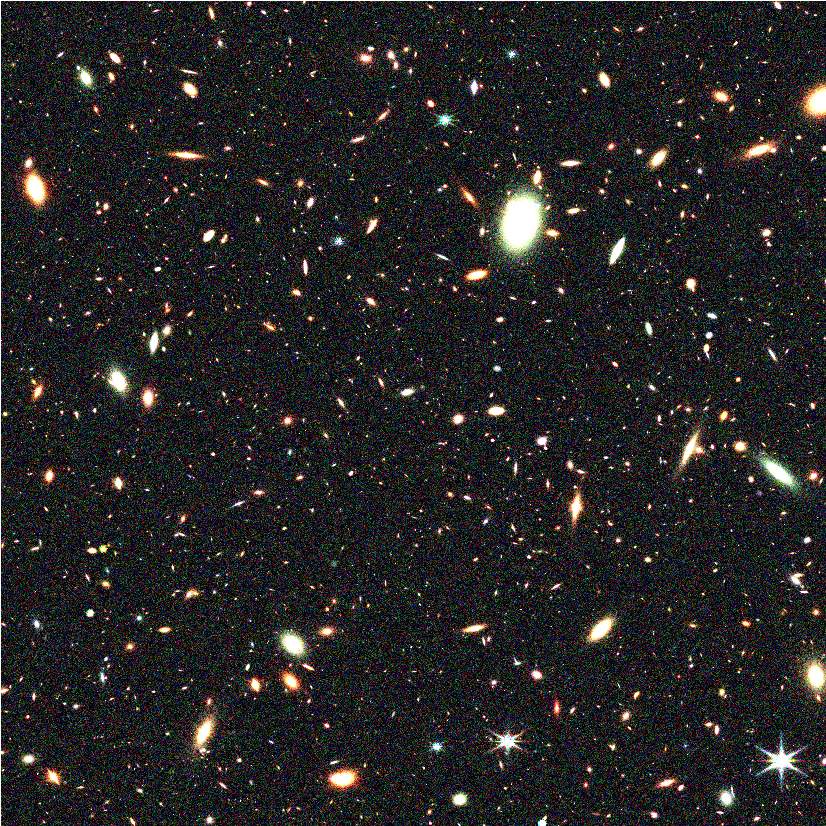}
	\caption{Simulated composite image in the NIRCam/F115W, F200W and F356W bands to a depth of $\sim29$~mag ($5\sigma$), following the CEERS observing strategy. The area is 4.5~arcmin$^2$ and the resolution 0.031"/pixel for all the images (see Section~\ref{section3.2}).}
	\label{ds9_composite_115_200_356_image}
\end{figure*}

Mock images are generated with SkyMaker \citep{bertin_skymaker:_2009} from the convolution of point-like and extended sources (with any Sersic index) with an external point spread function (PSF).
Modeled PSF images are created with webbpsf\footnote{\url{https://webbpsf.readthedocs.io/en/stable/}} 
for \textit{JWST} and TinyTim\footnote{\url{http://www.stsci.edu/software/tinytim/}} \citep{krist_20_2011} for \textit{HST}, with an oversampling of five to avoid aliasing effects. The G2V star spectrum from \citet*{castelli_is_2004} is taken as source to generate polychromatic PSFs.
In mock \textit{HST} images, we include an additional jitter (Gaussian blurring) tuned to recover the measured PSF full width at half maximum (FWHM) in real data. 
The modeled PSF files are multiplied by a radial Fermic-Dirac kernel to limit edge effects around bright sources.
Our noise model consists of a single (uncorrelated) Poisson component for photon noise. In real images we expect the noise to be sky-dominated especially for faint sources, we therefore neglect other noise components such as readout noise, inter-pixel capacitance and cosmic rays.
Background levels and detection limits can be estimated with the Exposure Time Calculators (ETC) for \textit{HST}\footnote{\url{http://etc.stsci.edu/}} and \textit{JWST}\footnote{\url{https://jwst.etc.stsci.edu/}} \citep{pontoppidan_pandeia:_2016}. We tune the background surface brightnesses to reproduce the predicted/measured depths in each band. 


In each of the considered observing strategies, we generate mock images of 11 $\times$ 11~arcmin$^2$ including all the sources from the mock catalogs. 
The resulting predictions are then scaled down by numbers to match the area of the planned observations. 
We generate all images directly at the NIRCam short-wavelength pixel scale of 0.031"/pixel (the smallest among the instruments), avoiding astrometric and resampling issues.
Saturation effects are neglected, since the effective saturation limit of stacked small exposure images, as well as the detector non-linearity, may be difficult to model.
Figure~\ref{ds9_composite_115_200_356_image} shows an example of a simulated composite image in three NIRCam bands.
Real images from future \textit{JWST} surveys may depart from our mock images because of neglected instrumental effects. 

\subsection{Source extraction}

Photometry is measured with SExtractor \citep*{bertin_sextractor:_1996} in the dual-image mode. We successively use the NIRCam/F115W, F150W and F200W images as detection image, then combine the extracted catalogs with a 0.2" matching radius. 
We use the "hot mode" SExtractor parameters from \citet{galametz_candels_2013} optimized for faint sources, and we checked that we can effectively recover the sources detectable by eye. The redshift distribution of the detected galaxies in the CEERS and the HUDF configurations are represented in Fig.~\ref{fig_Nz_input}. 
We do not mask bright sources. 

Aperture photometry generally provides the less noisy color measurements compared to Kron \citep{kron_photometry_1980} photometry MAG$\_$AUTO \citep{hildebrandt_cfhtlens:_2012}, however this requires the images to be PSF-matched. 
We compute the PSF-matching kernels to the \textit{HST}/F160W band PSF with pypher\footnote{\url{https://github.com/aboucaud/pypher}} \citep{boucaud_convolution_2016}, from the PSF files used to create the mock images (neglecting PSF reconstruction). Each initial PSF file is resampled to the pixel scale of the target PSF, then the kernel is computed, resampled to the image pixel scale and convoluted with the image \citep{aniano_common-resolution_2011}. 
Fluxes are measured in 0.5" diameter apertures \citep{mclure_new_2013,mcleod_new_2015,mcleod_z_2016}, ensuring at least 70\% point-source flux is included in all bands. 
The PSF FWHM (respectively the 80\% encircled energy radius for point source) are 0.145" (0.28") for the NIRCam 4.4~$\mu$m band and 0.25" (0.49") for the MIRI 7.7~$\mu$m band. 

\begin{figure}
	\centering
	\includegraphics[width=\hsize]{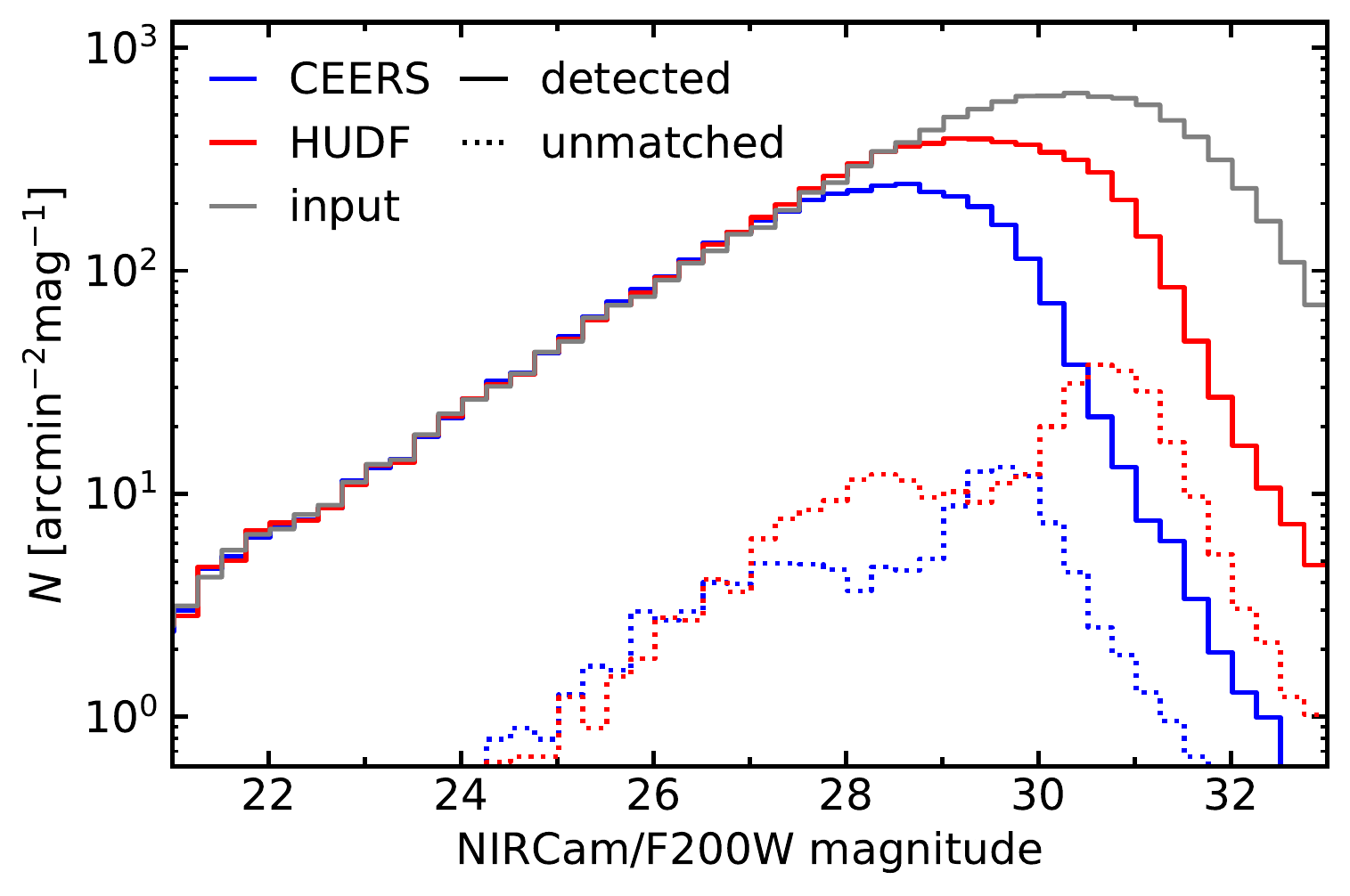}
	\caption{Detected source number counts versus NIRCam/F200W magnitude. The gray line indicates the input magnitudes of all the mock galaxies. The colored lines illustrate the measured magnitudes of detected sources in our CEERS and HUDF simulations. The solid lines include all the detected sources, the dashed lines represent unmatched sources only.
	}
	\label{fig_Nmag_input}
\end{figure}

Following \citet{laigle_cosmos2015_2016}, we apply corrections to the aperture photometry. SExtractor is known to underestimate flux errors in the case of correlated noise 
\citep[e.g.,][]{leauthaud_weak_2007}, arising from PSF-matching. We therefore apply a band-dependent correction to the measured flux errors, from the ratio of the median flux in empty apertures and the standard deviation of the source flux errors \citep{bielby_wircam_2012}. 
In addition, we scale both fluxes and flux errors with a source-dependent aperture to total correction, computed using MAG$\_$AUTO measurements \citep{moutard_vipers_2016}. 
%
We exclude truncated photometry and reject objects with negative aperture flux in all bands. 
Finally, we match the detected object positions with the input source catalog, taking the nearest match within a 0.1" search radius\footnote{Sources that are detected beyond this radius are either source pairs or false detections wrongly matched to undetected sources. The probability of the latter event is $\sim2$\%.}.
Figure~\ref{fig_Nmag_input} illustrates the detected number counts for both of the CEERS and HUDF configurations. The unmatched sources (indicated with dotted lines) present two components, the bright one including artifacts around stars and undeblended sources. We recall that the number of false detections is very sensitive to the noise model.

Galactic foreground extinction remains minimal in extragalactic fields at high galactic latitudes. In practice, it is often corrected by adjusting the image photometric zeropoints. 
We estimate the zeropoint corrections using the extinction curve of \citet{fitzpatrick_correcting_1999} and the Milky Way dust map from \citet*{schlafly_measuring_2011}. 
At the galactic latitude of the EGS field or the HUDF, the correction is at most 0.03~mag in the bluest considered band (ACS/F435W), therefore we decide to neglect galactic extinction in both the mock input spectra and the source extraction pipeline.

\subsection{Photometric redshift estimation}

To compute photometric redshifts, we perform SED-fitting with LePhare \citep{arnouts_measuring_2002,ilbert_accurate_2006}. Following \citet{ilbert_cosmos_2009}, we use the 31 templates including spiral and elliptical galaxies from \citet{polletta_spectral_2007} and a set of 12 templates of young blue star-forming galaxies using BC03 stellar population synthesis models. 
The BC03 templates are extended beyond 3~$\mu$m using the \citet{polletta_spectral_2007} templates, which include both PAH and hot dust emission from averaged \textit{Spitzer}/IRAC measurements. 
This set of templates has been extensively tested by the COSMOS collaboration \citep[e.g.,][]{onodera_deep_2012,laigle_cosmos2015_2016} and tested in hydrodynamical simulations \citep{laigle_horizon-agn_2019}. We do not include the two templates of elliptical galaxies added in \citet{ilbert_mass_2013} to avoid potential loss of information from degeneracies over the large redshift interval 
\citep{chevallard_modelling_2016}.
Dust reddening is added as a free parameter ($E(B-V )\le0.5$) and the following attenuation laws are considered: \citet{calzetti_dust_2000}, \citet{prevot_typical_1984}, and two modified Calzetti laws including the bump at 2175~\AA{} \citep*{fitzpatrick_analysis_1986}. 
Nebular emission lines are added following \citet{ilbert_cosmos_2009}. 
%
We impose that the absolute magnitudes satisfies $M_V>-24.5$ for CEERS and $M_B>-24$ in the HUDF, based on the LFs at $z<2$ in \citet{ilbert_vimos-vlt_2005} and assuming this is still valid at $z>2$. 
This SED-fitting prescription (e.g., SFH, attenuation) is distinct from the one used to generate the JAGUAR mock galaxies. This variability may reflect the potential disagreement between the fitted templates and reality, at least to a certain level.

The redshift probability distribution functions (PDF$z$) are measured in the redshift interval $0<z<15$. 
We perform SED-fitting using fluxes (not magnitudes) and do not use upper limits because this may remove essential information \citep{mortlock_probabilistic_2012}. 
We add a systematic error of 0.03~mag in quadrature to the extracted fluxes to include the uncertainties in the color-modeling (set of templates, attenuation curves). 
Photometric redshifts are defined as the median of the PDF$z$ \citep{ilbert_mass_2013}. 

Star templates are also fitted to reproduce and quantify potential object misclassification.
Similarly to \citet{davidzon_cosmos2015_2017}, we use the star templates from \citet{bixler_high_1991}, \citet{pickles_stellar_1998}, \citet{chabrier_evolutionary_2000}, the brown dwarfs templates from \citet{baraffe_new_2015} (see Sect.~\ref{Mock star sample}) and the BT-Settl grids with \citet{caffau_solar_2010} solar abundances at lower temperatures. These templates partly differ from the set of templates used to generate the mock stars.

We do not attempt to fit active galactic nuclei (AGN) templates. AGN-dominated SEDs typically present a featureless power-law optical-to-infrared continuum, strong emission lines and Ly$\alpha$ forest absorption especially at high redshifts. The observed emission of galaxies hosting AGNs strongly depends on the contribution of the two components. A large number of hybrid templates would be necessary to correctly characterize them, leading to risks of degeneracies in the SED-fitting procedure \citep{salvato_photometric_2009}.
In addition, AGNs exhibit variable emission with timescales from minutes to decades. Source variability may be observed from multiple-exposure imaging and dithering in both CEERS and the HUDF, so that AGNs brighter than the detection limit with relatively short timescales should be identifiable.

\subsection{Physical parameter estimation}

We run LePhare a second time following \citet{ilbert_evolution_2015} to determine other physical parameters such as stellar mass ($M_*$), star-formation rate\footnote{In LePhare, the measured SFR is instantaneous, whereas in JAGUAR it is averaged over the past 100~Myr. For exponential and delayed SFH with $\tau>1$~Gyr, the difference between these two SFR definitions is no more than 5\% (0.02~dex) at $z>0.1$.} 
(SFR) and absolute UV magnitudes ($M_\text{UV}$). Absolute magnitudes (uncorrected for attenuation) are computed using a top-hat filter of width 100~\AA{} centered at 1500~\AA{} rest-frame \citep{ilbert_vimos-vlt_2005}.
Redshifts are fixed to the photometric redshifts from the first LePhare run. 
The grid of fitted galaxy templates consists of BC03 models with exponential star-formation histories (SFH) with $0.1<\tau<30$~Gyr, and delayed SFH ($\tau^{-2}te^{-t/\tau}$) peaking after 1 and 3~Gyr. Two metallicities are considered ($Z_\odot$, $Z_\odot$/2). 
We allow $E(B-V)\leq0.5$ and only include the \citet{calzetti_dust_2000} starburst attenuation curve for simplicity and computational time (\citealt{ilbert_evolution_2015} included two attenuation curves).
Physical parameters are defined as the median of their marginalized probability distribution functions.


\section{Physical parameter recovery}
\label{section4}

\subsection{Photometric redshift recovery}
\label{section4_1}

\begin{figure*}[t]
	\centering
	\includegraphics[width=\hsize]{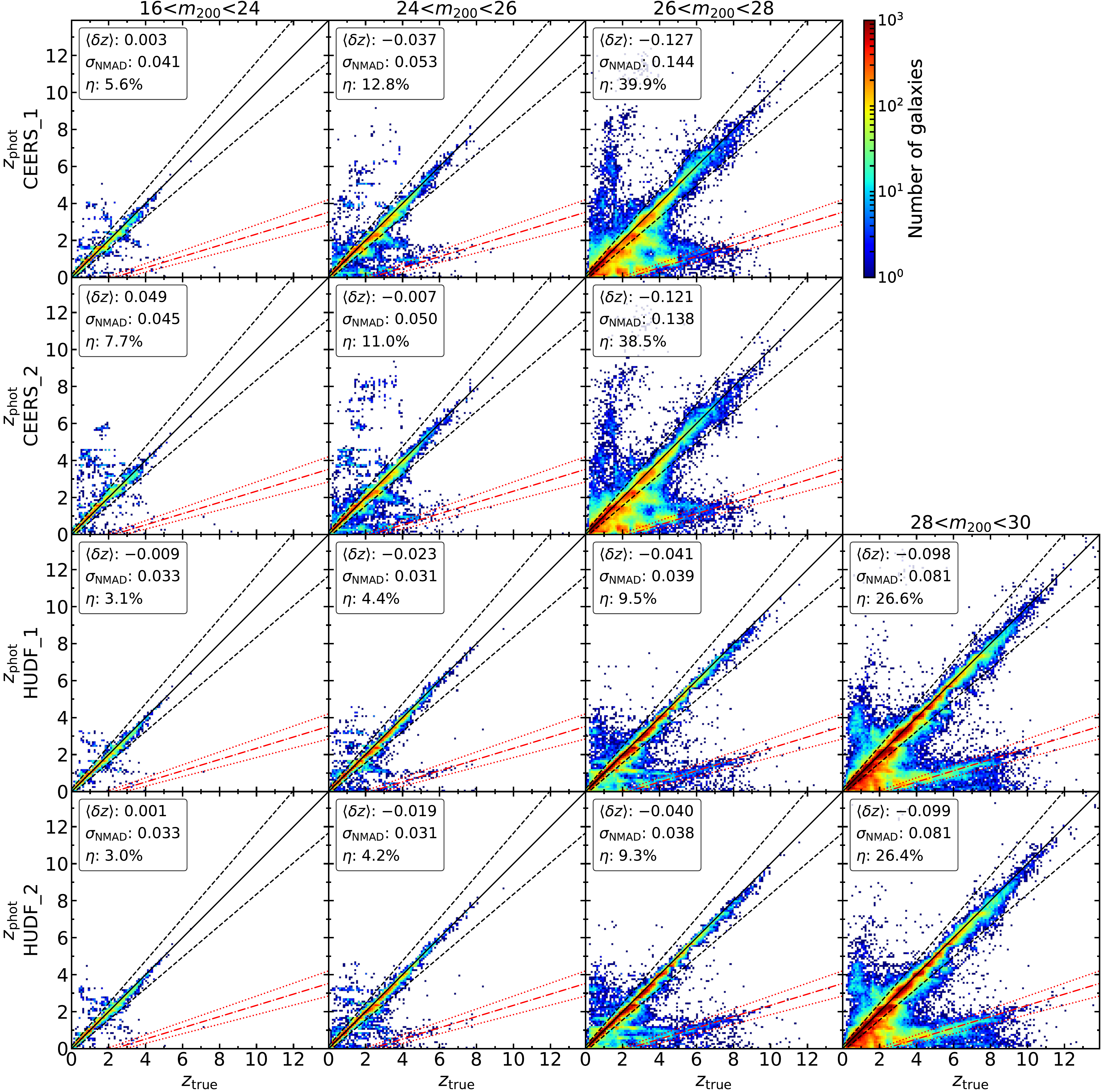}
	\caption{Comparison between photometric and true redshifts, "true" redshifts being in this case the simulated redshifts as described in Sect.~\ref{subsection_galaxy_sample}. The rows correspond to CEERS\_1, CEERS\_2, HUDF\_1 and HUDF\_2 observing strategies from top to bottom, and the columns represent observed NIRCam/F200W magnitude intervals. Color indicates the number of sources. The mean normalized residual, the normalized median absolute deviation and the catastrophic error fraction~$\eta$ for all the detected sources are indicated. The solid line shows the 1:1 relation and the dashed lines the $\pm0.15(1+z_\text{true})$ threshold used to compute $\eta$. The degeneracy between the Balmer 4000~\AA{} break and the Ly$\alpha$~1215~\AA{} break is identified by the red dotted-dashed line, with 15\% errors in dotted lines.}
	\label{fig_zphot_recovery}
\end{figure*}

The recovery of the photometric redshifts through SED-fitting can first be tested. The quality of the photometric redshifts is assessed with the following statistics \citep{ilbert_accurate_2006}:
\begin{itemize}
	\item the mean normalized residual $\langle\delta z\rangle$, with the normalized residuals $\delta z=(z_\text{phot}-z_\text{true})/(1+z_\text{true})$
	\item the normalized median absolute deviation (NMAD) $\sigma_\text{NMAD}=1.4826\times\text{med}(|\delta z-\text{med}(\delta z)|)$ 
	\item the fraction of catastrophic failures $\eta$, for which $|\delta z| > 0.15$
\end{itemize}

Figure~\ref{fig_zphot_recovery} represents the photometric and true redshifts for all the considered observing strategies, in multiple magnitude intervals. No selection is applied. 
%
%
We observe no systematic bias at $z_\text{true}<2$ in any configuration for the bright samples, for which the galaxy continuum redward the Balmer break is sufficiently well sampled.
However, the mean normalized residual becomes negative $\langle\delta z\rangle<-0.1$ at $z_\text{true}>2$, even in the brightest magnitude interval. 
%
This is probably due to the different attenuation curves in the mock galaxies and in LePhare. The effective, galaxy-wide attenuation curves of the JAGUAR mock galaxies (which employ the two-component attenuation law of \citealt*{charlot_simple_2000}) are typically grayer (flatter) than the \citet{calzetti_dust_2000} model in the infrared. 
The bump at 2175~\AA{} in the attenuation curve utilized in LePhare and not JAGUAR may also be an issue. 

In the CEERS\_1 observing strategies, the number of catastrophic failures is significant even in the brightest magnitude bin. There are several explanations for that.
At $z_\text{true}<4$, there is a significant number of sources whose redshift is underestimated. Attenuated blue galaxies may be confused with lower redshift unattenuated red galaxies. One of the main reasons for this is the degeneracy between the Lyman and the Balmer breaks, as confirmed from spectroscopic surveys \citep{le_fevre_vimos_2015}. This confusion is enhanced by the lack of optical data in the EGS field, with no deep imaging blueward \textit{HST}/F606W, so that the Balmer break cannot be correctly identified at low redshift. This is the main reason for the outliers among bright sources.
At $z_\text{true}>4$ the Ly$\alpha$ break becomes detectable in the \textit{HST} bands. The number of catastrophic redshift underestimates is therefore reduced, especially for bright sources thanks to the NIRCam bands sampling both the Balmer and Ly$\alpha$ breaks. 
Strong emission lines may lead to overestimating the continuum, especially for observing strategies that only employ broad-band filters. This can have a significant impact on determining the position of the Balmer break.
Quiescent galaxies appear to have a larger dispersion but a smaller outlier fraction than star-forming galaxies.

The two additional MIRI bands at 5.6 and 7.7~$\mu$m in the CEERS\_2 observing strategy 
marginally improve the photometric redshift estimates. Both dispersion and outlier rate are larger in the brightest magnitude interval and smaller at fainter magnitudes. 
At high redshift $z>4$, the MIRI filters cover the rest-frame near-IR or optical region, therefore sampling the stellar continuum or even the Balmer break. The photometric redshift dispersion is reduced by $\Delta\sigma_\text{NMAD}=0.01$ for $4<z_\text{true}<7$ galaxies. 
Most of the faint NIRCam-detected sources, however, are not detected in MIRI at the depths that will be probed by the CEERS survey. 
For low-redshift $z<4$ galaxies, the \textit{HST}+NIRCam bands impose most of the constraints on photometric redshifts. We still observe fewer catastrophic failures because of Lyman-break misidentification when MIRI data are available, and a systematic bias lowered by 0.05 at $z_\text{true}=2$. This comes from a improved sampling of the stellar continuum with MIRI. 
However, the number of outliers with $z_\text{true}<4$ and $z_\text{phot}>4$ at $m_{200}<26$ is increased.
%
One of the reasons for more outliers among bright sources with MIRI may be the treatment of dust. The key feature appears to be the observed-frame mid-IR colors. 
Galaxies with good photometric redshifts mostly present decreasing mid-IR emission with increasing wavelength, whereas outliers often present increasing mid-IR emission. This feature can appear in our mock galaxies from (1) large dust continuum, remaining non-negligible even at $\sim2-3$~$\mu$m rest-frame because of high dust temperature, or (2) large PAH emission lines at 3.3, 6.2 and 7.7~$\mu$m. The infrared luminosities may be overestimated, notably because of the energy balance assumption. In contrast, we are not performing an energy balance in the fitting with LePhare, so that the attenuation and dust emission are disconnected. 
In addition, the \citet{polletta_spectral_2007} templates include dust emission from averaged \textit{Spitzer}/IRAC measurements, so they may not include the mid-IR brightest galaxies. 
Consequently, LePhare tends to favor high-redshift solutions for low-redshift galaxies with bright and red mid-IR colors.
Because of these uncertainties in the mid-IR modeling, one could increase the systematic error added in quadrature to the MIRI photometry. However, this would reduce the additional mid-IR information that is essential to their characterization of high-redshift sources. We therefore do not follow this option.

The main improvements in the HUDF configurations are the deeper \textit{HST} and NIRCam photometry, leading to the considerable improvements in both the photometric redshift dispersion and outlier rates compared to CEERS. 
Spectral features such as the Lyman and Balmer breaks can be better captured with the twice more numerous \textit{HST} bands in the red and near-IR filters. As a consequence, the number of low-redshift galaxies at $z<3$ with $z_\text{phot}>4$ is significantly reduced. 
Moreover, the additional $B_{435}$ band offers an improved sampling of the Balmer break at $z<3$ and the Lyman break at $z>4$. We find that the global outlier rates and photometric redshift dispersion are decreased by about 10\% thanks to the addition of the blue band. 
Furthermore, the two NIRCam medium-bands marginally reduce the redshift outlier rates at $z>6$ mostly. 
%
With the additional MIRI/F560W band in HUDF\_2, we do not observe any improvement in the global photometric redshift dispersion or outlier rate. 
In contrast, both of them are improved at high redshift and especially at $z>10$, where MIRI provides the only information redward the Balmer break. 

\begin{figure}
	\centering
	\includegraphics[width=\hsize]{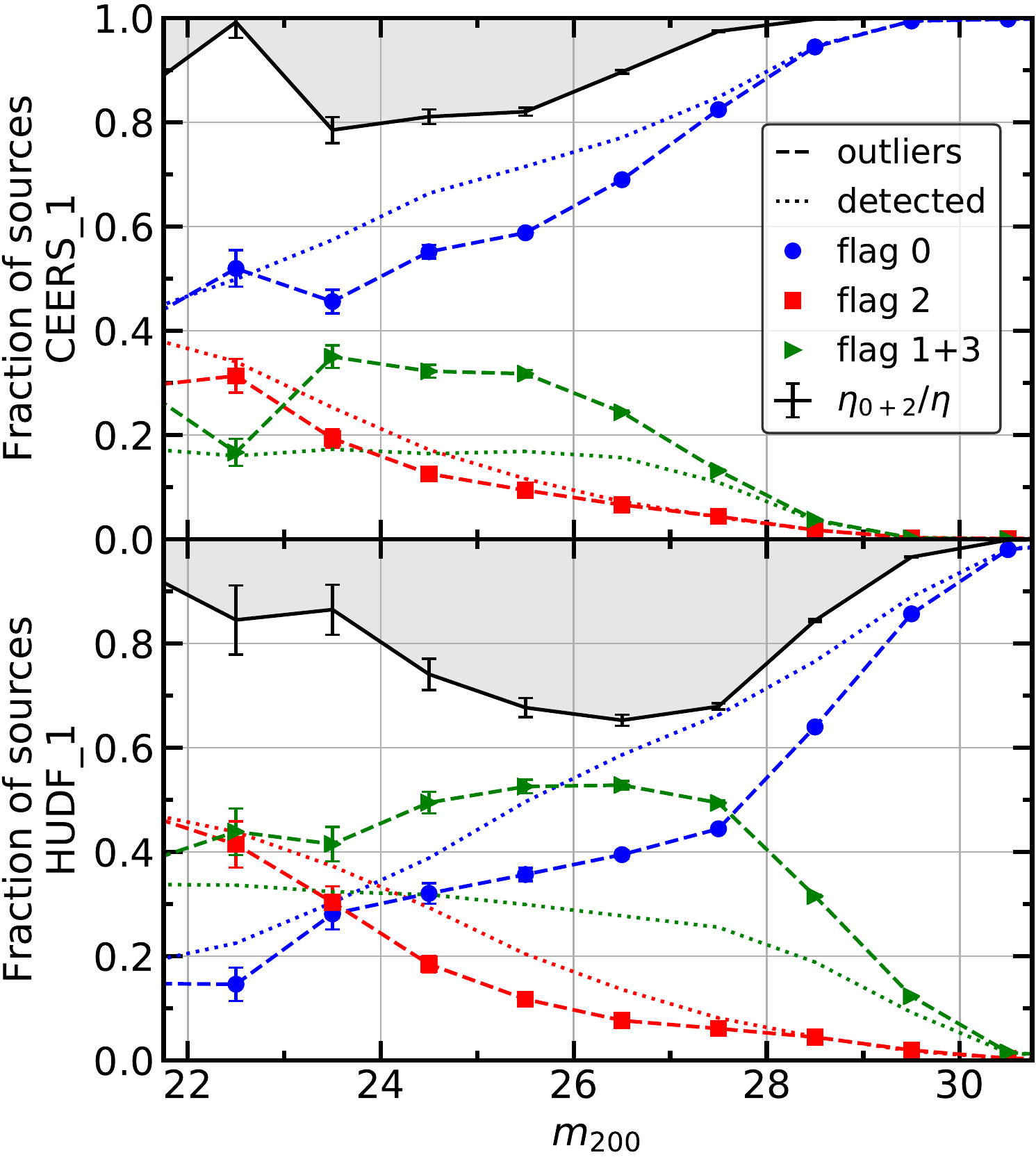}
	\caption{Fraction of detected sources per SExtractor flags indicating blending (flag 2), contaminated photometry (flag 1), both (flag 3) or none (flag 0) versus observed NIRCam/F200W magnitude. The rows correspond to the CEERS\_1 (top) and HUDF\_1 (bottom) observing strategies. The dashed lines indicate the flag fractions among photometric redshift outliers (summing to one), as defined in Sect.~\ref{section4_1}. The dotted lines represent the flag fractions in the whole detected sample (summing to one). The black solid line shows the ratio between the photometric redshift outlier rates $\eta_{0+2}$ assuming all the sources have uncontaminated photometry (flags 0 or 2), and the standard outlier rates. The error bars are propagated Poisson errors.}
	\label{fig_zphot_outlier_fraction}
\end{figure}

Source blending may also lead to catastrophic photometric redshifts, because of contaminated photometry and incorrect colors. 
The photometric redshifts of blended high-redshift galaxies tend to be mostly underestimated, which is coherent with blended source pairs or groups that are most likely to contain at least one galaxy at $z\sim1-2$. 
%
Figure~\ref{fig_zphot_outlier_fraction} illustrates the fraction of detected sources with the SExtractor flags indicating blending (flag 2), contaminated photometry (flag 1), both (flag 3) or none (flag 0). The solid lines represent the photometric redshift outliers and dotted lines the whole detected sample.
The number of flagged sources mostly decreases with increasing magnitudes, because faint sources typically need to be isolated to be detected, therefore unflagged. In contrast, faint sources with bright neighbors may remain undetected.
Flagged objects represent a large portion of the detected sources, about 40\% (65\%) at $m_{200}<26$ in CEERS (HUDF), which increases with the depth of the survey.
We observe a significantly increased fraction of contaminated sources (flags 1+3) in the $z_\text{phot}$ outliers, about twice as much as in the entire sample at $m_{200}<26$ in CEERS and at $m_{200}<28$ in the HUDF. 
%
In the hypothetical case where all the detected sources had uncontaminated photometry (flags 0+2), the photometric redshift outlier rates $\eta$ would be corrected by the indicated ratio $\eta_{0+2}/\eta$. This ratio reaches 80\% at $24<m_{200}<26$ in CEERS, meaning that the outlier rate would decrease from 12\% to 10\% in this magnitude bin. Similarly in the HUDF, the outlier rate at $26<m_{200}<28$ would decrease from 9\% to 6\%. 
We observe no significant wavelength dependence of these values. 
These results indicate that source blending will definitely be an issue with deep \textit{JWST} imaging.




\subsection{Stellar mass recovery}

\begin{figure*}[t]
	\centering
	\includegraphics[width=\hsize]{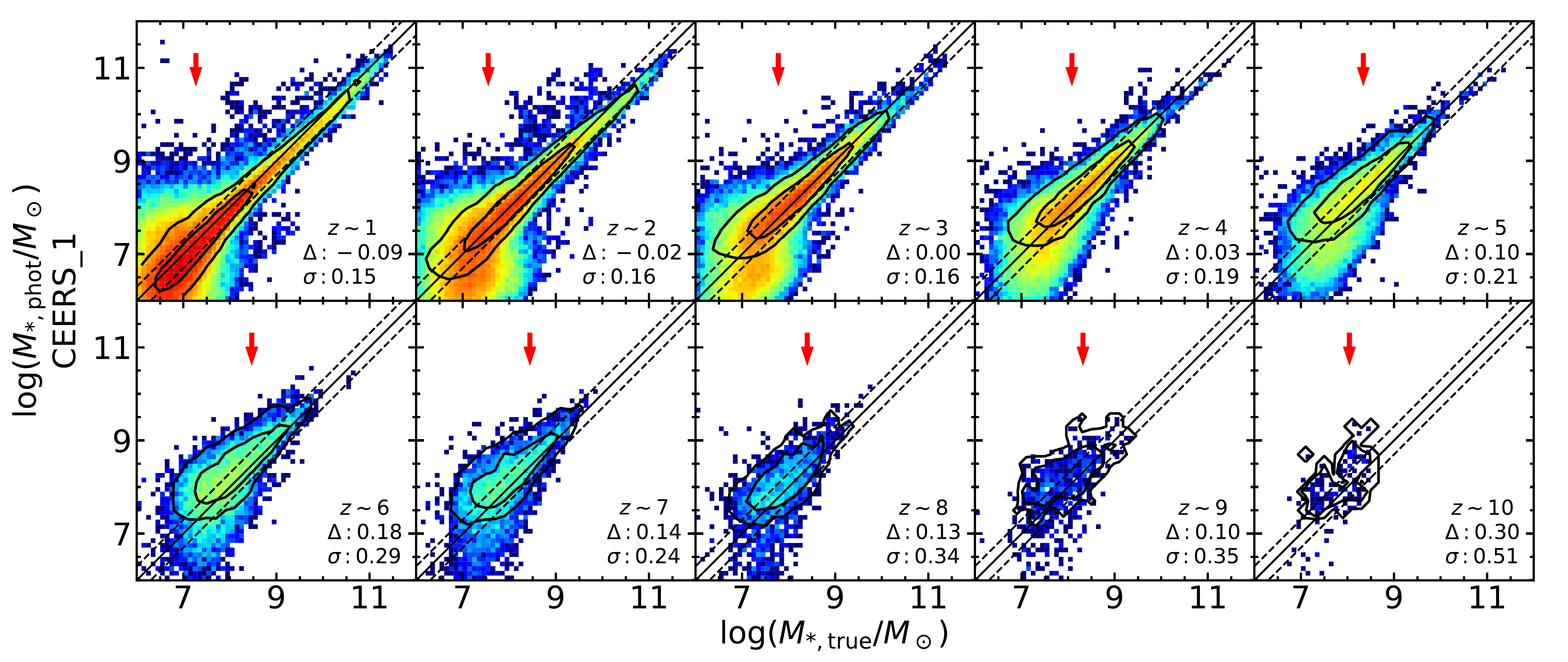}
	\includegraphics[width=\hsize]{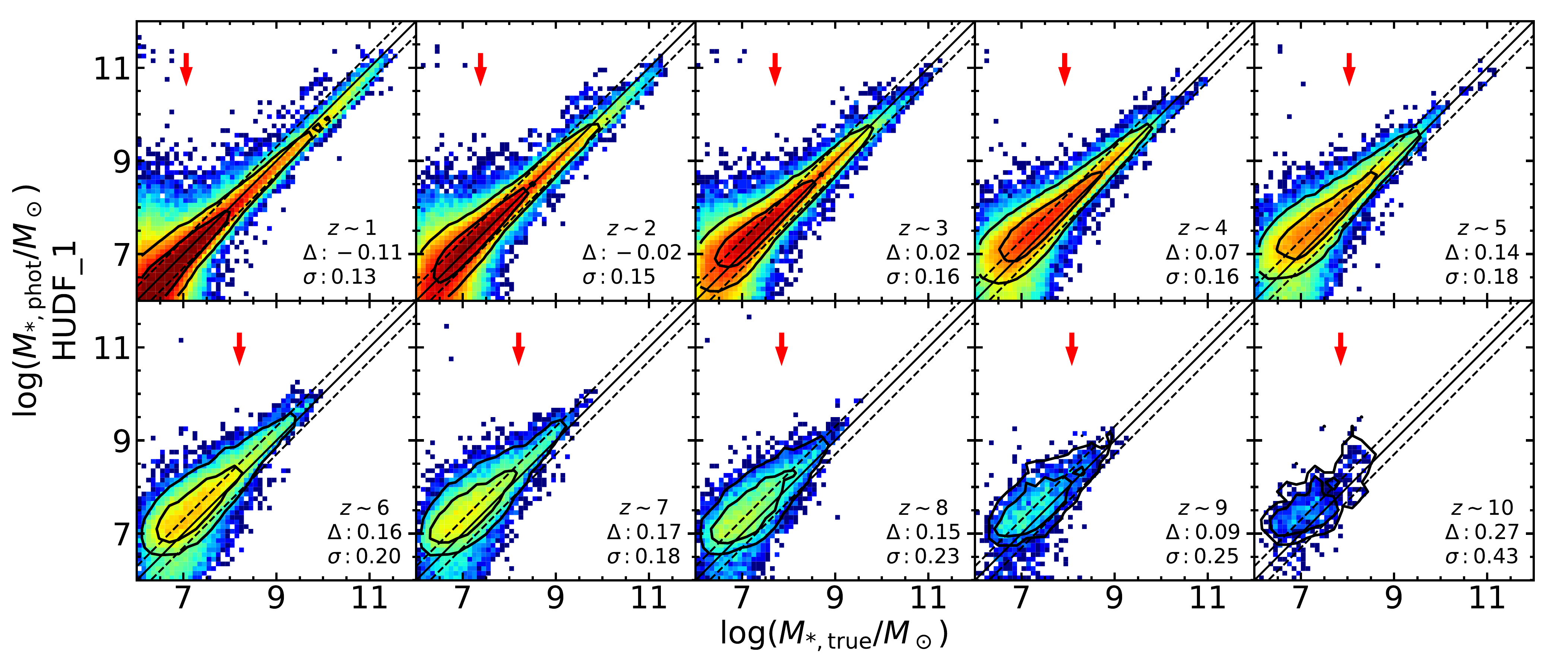}
	\caption{Comparison between measured and input stellar masses, for the CEERS\_1 and HUDF\_1 observing strategies (top and bottom figures, respectively). Each panel represents an input redshift interval centered at $z_\text{true}=1,2,3,...$ of width $\Delta z=1$. 
	Color indicates the number of sources in the whole sample, with the same color scale as in Fig.~\ref{fig_zphot_recovery}. The thick black contours represent the distribution of the sources with correct photometric redshift $|z_\text{phot}-z_\text{true}| < 0.15 z_\text{true}$, including 68\% and 95\% of these sources respectively. The median shift ($\Delta$) and the NMAD ($\sigma$) for the sources with correct $z_\text{phot}$ and $8<\log(M_\text{*,true}/M_\odot)<10$ are indicated (in dex). The solid line shows the 1:1 relation and the dashed lines $\pm0.3$~dex. The red arrows indicate the detected 90\% stellar mass completeness.}
	\label{fig_Mstar_recovery}
\end{figure*}

\begin{figure*}[t]
	\centering
	\includegraphics[width=\hsize]{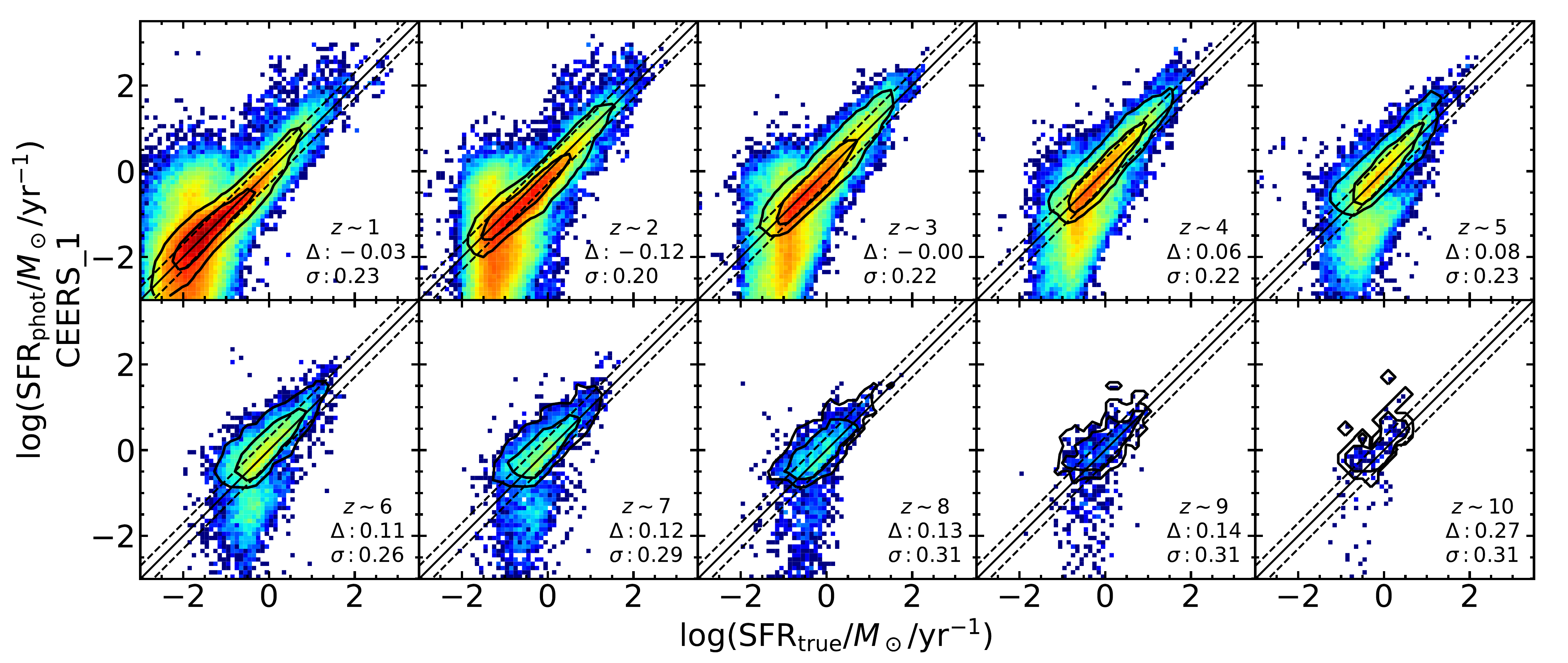}
	\includegraphics[width=\hsize]{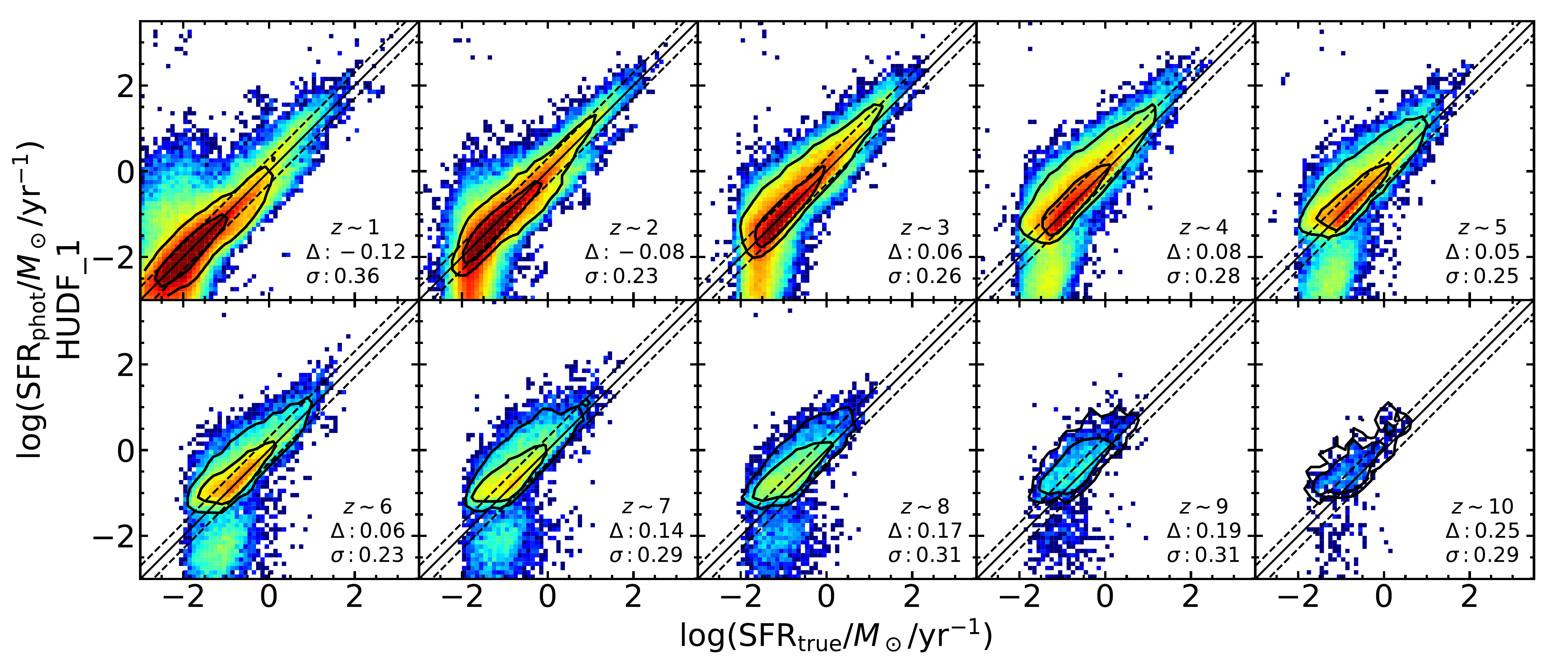}
	\caption{Same as Fig.~\ref{fig_Mstar_recovery} for star formation rate. The median shift ($\Delta$) and the NMAD ($\sigma$) for the sources with correct $z_\text{phot}$ and $-1<\log(\text{SFR}_\text{true}/M_\odot/\text{yr}^{-1})<1$ are indicated (in dex). The solid line shows the 1:1 relation and the dashed lines $\pm0.3$~dex.}
	\label{fig_SFR_recovery}
\end{figure*}

\begin{figure*}[t]
	\centering
	\includegraphics[width=\hsize]{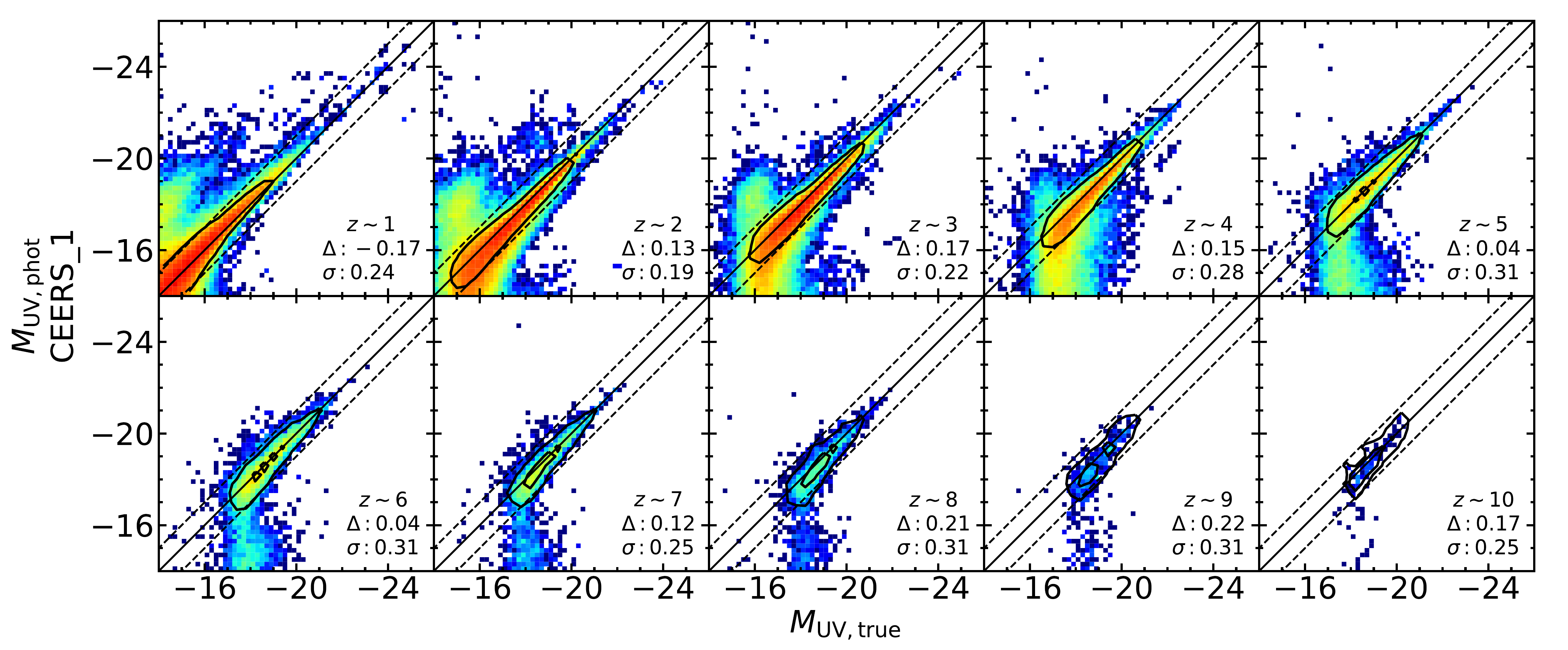}
	\includegraphics[width=\hsize]{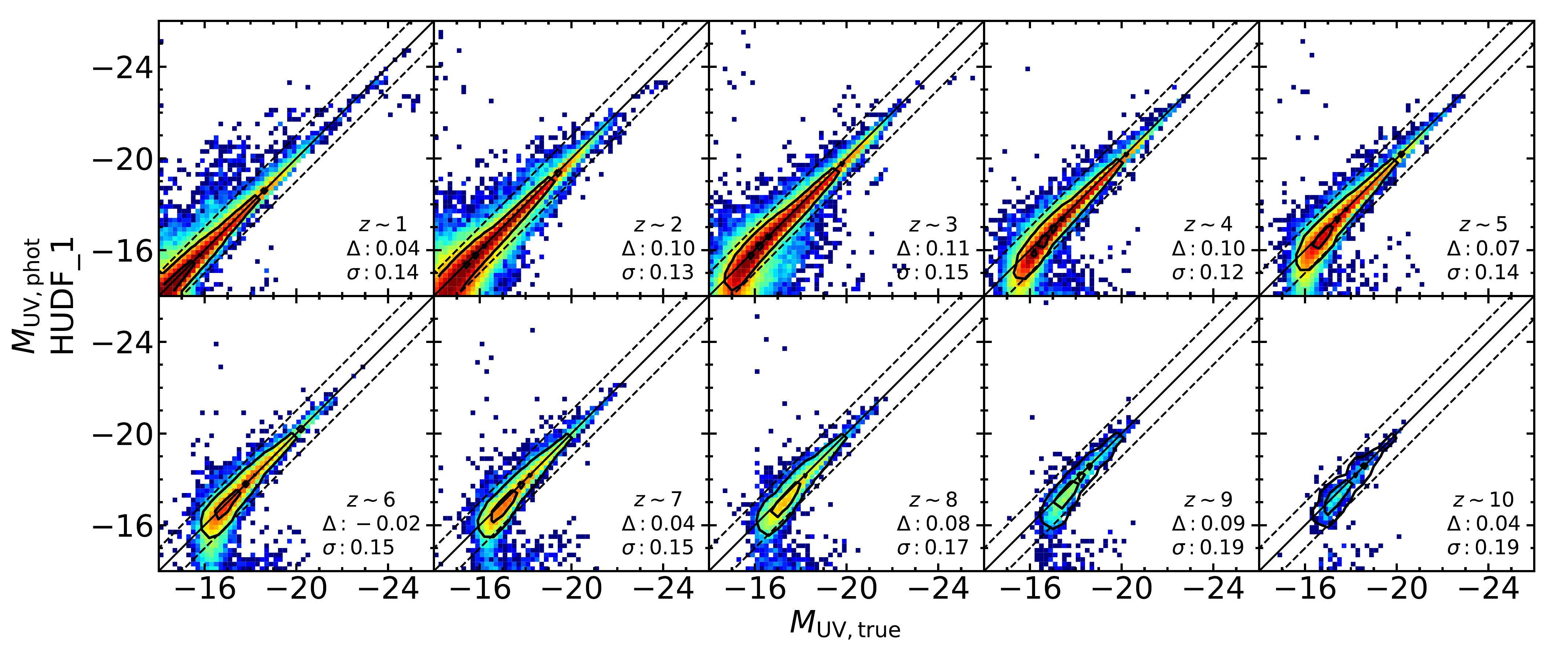}
	\caption{Same as Fig.~\ref{fig_Mstar_recovery} for absolute UV magnitude. The median shift ($\Delta$) and the NMAD ($\sigma$) for the sources with correct $z_\text{phot}$ and $-20<M_\text{UV}<-18$ are indicated (in mag). The solid line shows the 1:1 relation and the dashed lines $\pm1$~mag.}
	\label{fig_MUV_recovery}
\end{figure*}

The comparison between input stellar masses and those measured through SED-fitting is illustrated in Fig.~\ref{fig_Mstar_recovery} for the CEERS\_1 and HUDF\_1 observing strategies. The redshift intervals are centered at $z=1,2,3,...$ with a width of $\Delta z=1$.
The measured stellar masses agree well with the input ones. 

In the CEERS\_1 configuration, the stellar mass dispersion is below 0.25~dex at $\log(M_*/M_\odot)>9$. Removing the photometric redshift outliers lowers the dispersion at $\log(M_*/M_\odot)>9$ to 0.2~dex, and significantly reduces the number of catastrophic stellar mass estimates at $\log(M_*/M_\odot)<8.5$. These low-mass objects are typically fainter and have noisier colors. The overall dispersion increases as the stellar mass decreases, from 0.25~dex to 0.5~dex for $\log(M_*/M_\odot)$ of 9 and 8 respectively, and for galaxies with good photometric redshifts, from 0.2~dex to up to 0.35~dex.
Most of the outliers at $z<4$ have stellar masses that are overestimated because of overestimated redshifts. The remaining cases are galaxies with nearby sources, boosting their aperture fluxes and affecting their colors. For blended galaxies with nevertheless correct colors and photometric redshifts, the total fluxes may not be correctly recovered through the deblending procedure despite the aperture-to-total flux correction.
In the HUDF\_1 strategy, both the number of outliers and the dispersion are smaller for low-mass galaxies, remaining below 0.45~dex up to $\log(M_*/M_\odot)=7$ and even below 0.35~dex when discarding catastrophic photometric redshifts. Stellar masses are not significantly affected by redshift outliers above $\log(M_*/M_\odot)>8$. The dispersion remains below 0.2~dex at $\log(M_*/M_\odot)>9$ at all redshifts. These results mainly reflect the improvements in the photometric redshifts from deeper \textit{HST} and NIRCam imaging and with the additional \textit{HST} blue band. Moreover, the near-IR medium-band photometry enable the emission lines and the galaxy continuum to be better separated. The systematic overestimation and the dispersion at $z\geq6$ are lowered by 0.05~dex only thanks to the medium bands, which are located close to the Balmer break in the rest-frame.

The median stellar mass lies between $\pm0.2$~dex around the input value between $8<\log(M_*/M_\odot)<10$ at all redshifts and in both configurations, and is generally underestimated at $z\leq3$ and overestimated at $z>4$. These observations may again come from the steepness at $\lambda>1$~$\mu$m of the attenuation curves used in input and in the SED-fitting (see Sect.~\ref{section4_1}). More attenuation may hide more low-mass stars and therefore result in underestimated mass. 
At $\log(M_*/M_\odot)<8$, the stellar mass estimates are systematically overestimated for galaxies with correct redshifts, the bias increasing with redshift and by up to 0.6~dex/$\log(M_*/M_\odot)$. 
At $\log(M_*/M_\odot)>10$ and $z<6$, stellar masses are systematically underestimated by $0.15-0.2$~dex. Massive galaxies are typically the most attenuated, and these galaxies effectively have large input attenuation $\hat{\tau}_V>0.1$. The percentage of $\log(M_*/M_\odot)>10$ galaxies with $\hat{\tau}_V>1$ is 57\%, and reaches 70\% for the subset where mass is underestimated by at least 0.2~dex. Strong attenuation $E(B-V)>0.5$ ($A_V>2$) are not allowed in our LePhare configuration to avoid additional degeneracies between templates. The underestimated attenuation in SED-fitting may lead to underestimated stellar mass. 
%
In contrast, galaxies at $\log(M_*/M_\odot)\sim9$ with $\hat{\tau}_V>0.2$ have overestimated stellar masses, by 0.1, 0.2, 0.3~dex at $z=4,5,6$ in both CEERS and the HUDF.

Quiescent galaxies have underestimated stellar masses in all the observing strategies, by 0.15~dex at $\log(M_*/M_\odot)>9$ and by 0.5~dex below. These numbers are not reduced when removing photometric redshift outliers. High-mass quiescent galaxies typically have large metallicity, however observational constraints on the metallicity of low-mass galaxies are lacking. In JAGUAR, low-mass ($\log(M_*/M_\odot)<8.7+0.4z$) quiescent galaxies are assigned random uniform metallicities between $-2.2<\log(Z/Z_\odot)<0.24$. The recovered stellar masses of $\log(M_*/M_\odot)<9$ quiescent galaxies with $\log(Z/Z_\odot)<-0.5$ ($>-0.5$) are underestimated by up to 0.7 (0.4)~dex.
This dramatic underestimation of stellar mass for low-mass quiescent galaxies may come from the quiescent galaxy templates in LePhare that do not span the parameter space of the mock galaxies. In particular, only two metallicities ($\log(Z/Z_\odot)=0,-0.3$) are allowed in the LePhare configuration to avoid degeneracies between templates. 
In addition, dust attenuation was neglected for low-mass quiescent galaxies in JAGUAR, which may also explain this systematic bias at low masses.

The CEERS\_2 strategy presents results equivalent to CEERS\_1. This is not surprising because of the galaxy continuum already well sampled with NIRCam. At very high redshift $z>10$ where NIRCam does not sample redward of the Balmer break, the photometric redshift estimates still rely on NIRCam, and the shallow MIRI imaging did not significantly improve stellar mass estimates. 
%
With HUDF\_2 however, the MIRI data are deep enough to slightly improve stellar masses at $z\geq9$, with the scatter and systematic bias lowered by about 0.05~dex. This essentially comes from the improvement of photometric redshifts.

\subsection{Star formation rate recovery}

Figure~\ref{fig_SFR_recovery} illustrates the galaxy star formation rate recovery in the CEERS\_1 and HUDF\_1 observing strategies. The results with the CEERS\_2 and HUDF\_2 configurations, respectively, are strictly similar. 

The measured SFRs remain in correct agreement with the input values, however less precise than stellar mass estimates. 
We note that the SFR estimates may behave well because the assumed SFH in LePhare and in JAGUAR are similarly simple, meaning smooth exponential or delayed SFH. 
The low precision is primarily due to the degeneracy between SFR and dust attenuation, which affects the rest-frame UV, where the emission is dominated by hot, young stars. 
In an analogous work, \citet{laigle_horizon-agn_2019} showed that with a similar LePhare configuration, attenuation is the main source of systematic uncertainties and dispersion in the SFR recovery. 
In addition, the missing nebular continuum emission in LePhare may also be an issue.
%
For galaxies with good photometric redshifts, the SFR dispersion is 0.3~dex for the CEERS survey and 0.35~dex for HUDF, and remains stable over redshift and input SFR. 
In the HUDF, however, the recovered SFR distributions are skewed toward large SFRs at $z\geq3$. 
This surely comes from the more difficult match between the input and the fitted galaxy templates, because of the increased depth in the HUDF and, at low redshift, the \textit{HST} $B$-band, giving stronger constraints on the SFR tracers. 

We observe that the median shift at SFR$_\text{true}>1$~$M_\odot/$yr is bounded by $\pm0.2$~dex at all redshifts in all the observing strategies.
In particular, the most star-forming galaxies at $z<6$ with SFR$_\text{true}>10$~$M_\odot/$yr have systematically overestimated SFR estimates by 0.15~dex. This may come from the difference of attenuation curves assumed in JAGUAR and in LePhare. 
Additionally, most of the outliers at SFR$_\text{true}>1$~$M_\odot/$yr have overestimated SFR estimates, similarly to the stellar mass outliers. 
Among galaxies with correct photometric redshifts, the systematic bias increases with decreasing SFR$_\text{true}$ and as redshift increases, reaching 0.5~dex at $0.1$~$M_\odot/$yr in CEERS and 0.4~dex in the HUDF.
%
The large number of catastrophic failures at SFR$_\text{true}<1$~$M_\odot/$yr at all redshifts comes from redshift misestimation. This feature appears in all the observing strategies, and its importance is only slightly reduced in the HUDF compared to CEERS. In our methodology to estimate galaxy physical parameters, imposing underestimated redshifts in the second SED-fitting run gives underestimated SFRs and vice-versa. This would be a priori unknown in real surveys, so it shows the importance of simulations to make the necessary corrections.


\subsection{Absolute UV magnitude recovery}
\label{section_UV_recovery}

Figure~\ref{fig_MUV_recovery} illustrates the recovery of absolute UV magnitudes. There are features in common with the stellar mass and SFR measurements, such as outliers with mostly overestimated luminosities, and the dispersion from catastrophic photometric redshifts.
%
In the CEERS configurations, the dispersion increases from 0.2~mag at $M_\text{UV}=-20$ to 0.3~mag at $M_\text{UV}=-18$ for $z\leq3$ galaxies. At higher redshift, the distributions are typically 0.1~mag broader. 
The UV luminosities are overestimated by 0.15~mag at $z\sim1$ and underestimated by at most 0.2~mag at $z\geq2$ for sources with $M_\text{UV}>-18$ and good photometric redshifts. 
%
In comparison, in the HUDF, the dispersion at $M_\text{UV}<-18$ remains below 0.2~mag at all redshifts for sources with correct photometric redshifts, and below 0.25~mag (0.4) at $M_\text{UV}<-17$ and $z\leq3$ ($\geq3$). The magnitudes are systematically overestimated by $\sim0.1$~mag at $M_\text{UV}>-18$. For low-redshift $z<2$ galaxies, the improvements in the HUDF are driven by the additional $B_{435}$-band photometry and the smaller K-corrections required to compute the absolute UV magnitudes. 
In the JAGUAR galaxies, the birth cloud component of the dust attenuation may strongly affect the rest-frame UV emission. LePhare may underestimate the attenuation especially at this wavelength, leading to underestimated UV luminosities. 
The NIRCam medium bands decrease the systematic bias and dispersion by about 0.03~mag at $z\ge6$.

\subsection{Comparison with previous works}

\citet{bisigello_impact_2016,bisigello_recovering_2017,bisigello_statistical_2019} investigated the recovery of the galaxy photometric redshifts and physical parameters with \textit{JWST} broad-band imaging. 
The authors considered multiple galaxy samples, observed galaxies at $z<7$ and simulated spectra constructed from BC03 and \citet{zackrisson_spectral_2011} population synthesis models at $z>7$.
All the combinations of a few discrete physical parameters were used to build the high-redshift galaxy samples. 
As a consequence, the distribution of these parameters among the real galaxy population at the given redshift was not respected. 
In addition, the source samples did not reproduce the redshift distribution of a flux-limited galaxy population, meaning that the contamination from foreground low-redshift galaxies into the high-redshift samples could not be estimated.
The galaxy physical properties were then determined through SED-fitting with LePhare using the same galaxy templates as for the input spectra. Stellar and brown dwarf templates were not fitted, meaning that the nature of the sources was assumed to be known a priori.
\citet{bisigello_impact_2016} already showed that \textit{HST} short-wavelength optical data could significantly reduce the photometric redshift dispersion and outlier rate.
\citet{bisigello_recovering_2017} notably investigated the stellar mass recovery for $7<z<10$ galaxies with the eight NIRCam broad-bands and MIRI imaging. The recovered precision on stellar masses were similar to our results, as well as the systematic overestimation attributed to emission lines.

\citet{kemp_maximizing_2019} analyzed the redshift and stellar mass recovery with \textit{JWST} and \textit{HST} imaging. The authors notably used the Empirical Galaxy Generator (EGG, \citealt{schreiber_egg:_2017}) to generate a complete magnitude-limited sample of $0<z<15$ and $5<\log(M_*/M_\odot)<12$ galaxies over 1.2~deg$^2$. This catalog included individual spectra with no emission lines, nonetheless the case of emission lines was treated with another sample of mock galaxies. 
The authors introduced and investigated two observing strategies including eight NIRCam bands with MIRI/F770W parallels, and \textit{HST}/$V_{606}$ and $i_{814}$ bands as ancillary data. These configurations are similar to the CEERS program, with similar choices of filters, exposure times and depths in the deepest regions. 
We come to the similar conclusions about MIRI, namely that its addition leads to an improvement in the photometric redshift recovery at $4<z<7$, though most of the constraints are coming from NIRCam and \textit{HST}. The authors quantified the gain from additional deep \textit{HST}/$B_{435}$ imaging, which was revealed to be more important than whether MIRI imaging was available. 

\section{Source selection}
\label{section5}

In this section, we investigate the selection of high-redshift galaxies and the rejection of contaminants, then we give predictions about the number counts and the recovery of the galaxy luminosity function. 
The impact of the selection on the galaxy samples is assessed using galaxy completeness and purity. We define completeness as the fraction of input galaxies, in a given magnitude $m$ and redshift $z$ bin, that are selected and assigned to the correct redshift interval. 
Likewise, purity is the fraction of selected sources, in an observed magnitude $m$ and redshift $z$ bin, that are high-redshift galaxies in this redshift interval. 
We define the redshift intervals $[z_i\pm\Delta z/2]$ with a width of $\Delta z=1$ and centered at $z_i=1,2,3,...$ Let $N_\text{input}$ be the number of input galaxies, $N_\text{selected}$ the number of selected sources assigned to a redshift bin, and $N_\text{correct}$ the number of selected galaxies that are assigned to the correct redshift interval. $N_\text{selected}$ may include false detections. Completeness $C$ and purity $P$ can be written as: 
\begin{equation}
C(m,z) = \dfrac{N_\text{correct}(m,z)}{N_\text{input}(m,z)},
\end{equation}
\begin{equation}
P(m^\text{obs},z) = \dfrac{N_\text{correct}(m^\text{obs},z)}{N_\text{selected}(m^\text{obs},z)}.
\end{equation}

The number of detected objects depends on the observation and the source extraction, setting the maximum number of sources that can be recovered. There is then a trade-off between completeness, purity and sample size: no selection will give maximum completeness (maximum sample size) and likely minimum purity, whereas stringent selections will lower completeness (lowering sample size) and likely higher purity. 

\subsection{Star rejection}
\label{Star_rejection}

\begin{figure}[t]
	\centering
	\includegraphics[width=0.8\hsize]{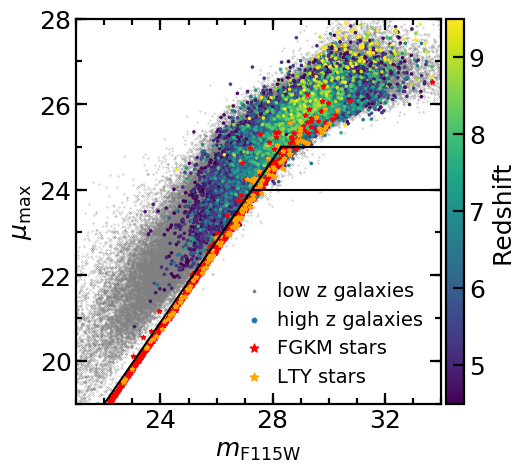}
	\caption{Maximum surface brightness-magnitude selection criteria to remove stellar objects. Each marker represents the measured colors of a detected source in the CEERS\_1 observing strategy. The colored points are high-redshift galaxies and the gray points indicate $z<4.5$ galaxies. The red and orange stars represent FGKM and LTY stars, respectively. 
	}
	\label{fig_stellar_mumax}
\end{figure}

\begin{figure}[t]
	\centering
	\includegraphics[width=0.8\hsize]{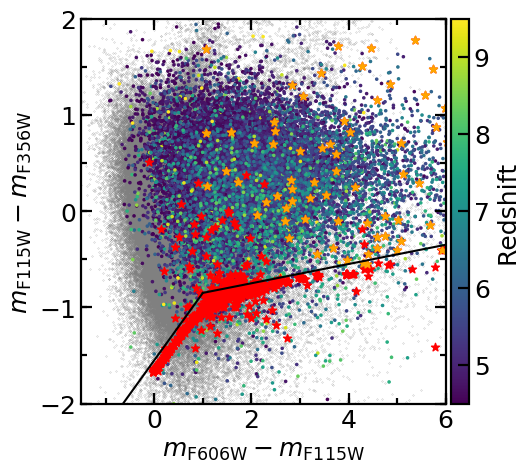}
	\includegraphics[width=0.8\hsize]{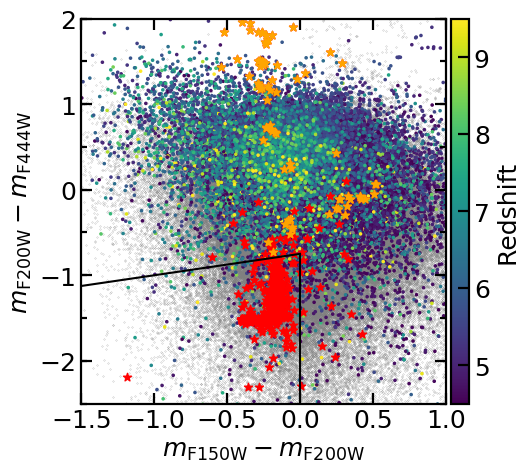}
	\caption{Color-color selection criteria to remove stellar objects. Each marker represents the measured colors of a detected source in the CEERS\_1 observing strategy. The colored points are high-redshift galaxies and the gray points indicate $z<4.5$ galaxies. The red and orange stars represent FGKM and LTY stars, respectively.
	Only sources detected at $2\sigma$ in the two reddest bands (in each panel) are indicated. 
	}
	\label{fig_stellar_color_color}
\end{figure}

\begin{figure*}[t]
	\centering
	\includegraphics[width=\hsize]{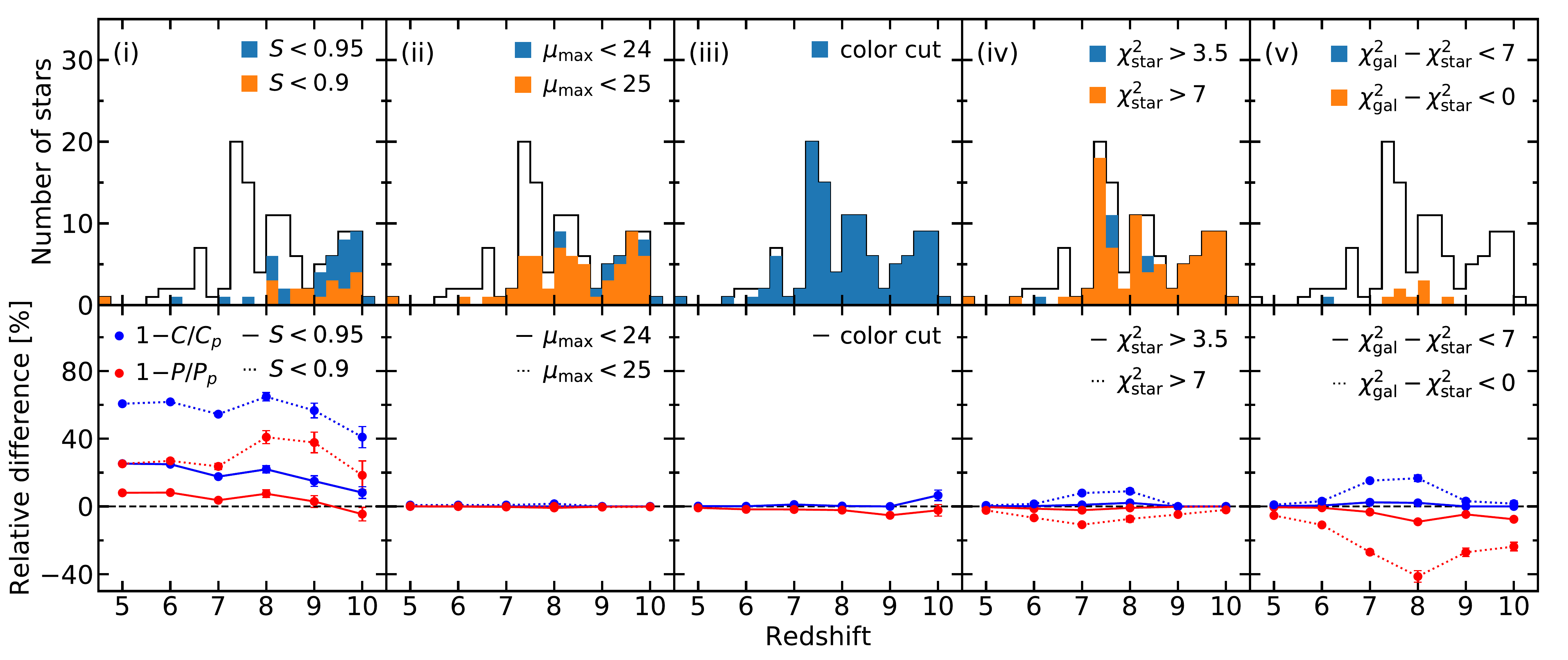}
	\caption{\textit{Top panels:} Number of remaining stellar contaminants after applying the indicated selection criterion per photometric redshift interval, in the CEERS\_1 observing strategy. Each column corresponds to one type of selection from Sect.~\ref{Star_rejection}. The black line indicates the photometric redshift distribution of the detected stars with $z_\text{phot}>5$. All the stars with lower photometric redshifts have $z_\text{phot}<2$.  
	The colored bars represent the remaining stars after applying the indicated selection criterion. The soft selection (in blue) is always less restrictive than the stringent one (in orange), meaning that all second counts are also included in the first counts.
	\textit{Bottom panels:} Completeness $C$ (in blue) and purity $P$ (in red) of the high-redshift galaxy samples versus true redshift (integrated over magnitudes). The references $C_p$ and $P_p$ represent the completeness and purity assuming the selection is based on photometric redshifts only. The relative difference with respect to the reference is represented, so that positive values indicate lower completeness and purity. 
	Different line styles represent the results for different selection criteria (as indicated in each panel).}
	\label{fig_stellar_criteria}
\end{figure*}

We first investigate the rejection of stellar objects contaminating the high-redshift galaxy samples. Commonly used criteria rely either on magnitude, colors, shape (or surface brightness) and the quality of the SED-fitting. We consider the following list of standard star rejection criteria, and we investigate the individual impact of each of them:
\begin{enumerate}[(i)]
	\setlength\itemsep{0.2em}
	\item $S<k$, with $k=0.95,0.9$
	\item $(\mu_\text{max}>0.95m_\text{115}-1.9)$ $\vee$ $(\mu_\text{max}>k)$ in CEERS, \\ $(\mu_\text{max}>0.95m_\text{115}-2.2)$ $\vee$ $(\mu_\text{max}>k)$ in the HUDF, with $k=24,25$
	\item $(m_\text{115}-m_\text{356} > 0.7(m_\text{606}-m_\text{115})-1.55)$ $\vee$ $(m_\text{115}-m_\text{356} > 0.1(m_\text{606}-m_\text{115})-0.95)$ if SNR$_\text{606}>2$, $(m_\text{150}-m_\text{200} > 0.25(m_\text{200}-m_\text{444})-0.75)$ $\vee$ $(m_\text{150}-m_\text{200} > 0)$ otherwise
	\item $\chi^2_\text{star} > k$, with $k=\nu/2,\nu$
	\item $\chi^2_\text{gal}-\chi^2_\text{star} < k$, with $k=\nu,0$
\end{enumerate}
with $S$ defined as the source stellarity index, $\mu_\text{max}$ is the maximum surface brightness, $\chi^2_\text{star}$ and $\chi^2_\text{gal}$ are the mean squared error from the SED-fitting of stellar and galaxy templates respectively, $\nu$ is the number of degrees of freedom in the fitting (set to the number of bands minus three). The thresholds $k$ define a soft (first value) and a stringent (second value) version for some selections. 

The first criterion (i) is based on the stellarity index $S$ measured with SExtractor in the NIRCam/F200W detection image. This is the posterior probability of a detected object to be a point-source (0 for extended source, 1 for point-source), according to its surface brightness profile. 
With high resolution imaging, brown dwarfs may be separated from resolved galaxies based on size \citep{tilvi_discovery_2013}. However, distant galaxies commonly appear point-like (e.g., bright star-forming blobs, faint galaxy hosting a bright AGN) and should not be discarded. The impact of this selection on galaxy completeness therefore depends on the morphology of the galaxies. This will need to be further investigated with simulations of more realistic galaxy light distributions. 
Similarly, stars tend to occupy a tight locus in the size (or surface brightness) - magnitude plane \citep{leauthaud_weak_2007}. We construct the selection (ii) in the maximum surface brightness $\mu_\text{max}$-$m_\text{115}$ plane as represented in Fig.~\ref{fig_stellar_mumax}. The parameter $\mu_\text{max}$ is the surface brightness [mag/arcsec$^2$] of the brightest pixel belonging to the source, above the estimated background. The NIRCam/F115W band is well adapted since stars mainly become fainter in redder bands and the emission of MLTY dwarfs drops in bluer bands.
We then make use of the color-color selections (iii) following \citet{davidzon_cosmos2015_2017}. The adopted color diagrams are (\textit{HST}/F606W-NIRCam/F115W) vs (NIRCam/F115W-F356W) for objects detected at 2$\sigma$ in the \textit{HST}/F606W band, and (NIRCam/F150W-F200W) vs (NIRCam/F200W-F444W) for the other sources (Fig.~\ref{fig_stellar_color_color}). 
%
Finally, we consider selections based on the SED-fitting results, either (iv) the absolute quality of the stellar fit \citep{bowler_galaxy_2015}, or (v) the relative quality of the stellar fit with respect to the galaxy fit \citep{ilbert_cosmos_2009}. 

Figure~\ref{fig_stellar_criteria} illustrates the photometric redshift distribution of stellar objects in the CEERS\_1 configuration, and the number of remaining stars after each rejection criterion is individually applied. In addition, the resulting differences of purity and completeness for the galaxy samples are indicated, for each redshift interval and integrated over magnitude. The purpose is to remove as many stellar contaminants as possible while maintaining a high galaxy completeness, and any gain in galaxy purity is an additional advantage in terms of statistics of the recovered galaxy population. 
As expected, the stellarity index cuts (i) manage to efficiently reject stars, about 65\% (80\%) for $S<0.95$ (0.9), however lowering galaxy completeness of about 20\% (60\%) at all redshifts $z>4$. In contrast, the surface brightness-magnitude selections (ii) remove a similar number of stars and maintain a high completeness and purity. Again, these impact on galaxy completeness depends on the assumed galaxy morphologies.
The color-color cuts (iii) have a marginal effect on brown dwarfs with $z_\text{phot}>5$, whereas most of the stellar objects with $z_\text{phot}<2$ are effectively removed. Neither galaxy completeness nor purity are much affected. The optical and near-IR colors of cold brown dwarfs appear not to occupy the same stellar locus as hotter stars, and removing them in the color-color space would discard many galaxies at the same time.
%
Finally, the criteria based on the absolute quality of the stellar fit (iv) only reject about 30\% of the stars, though slightly modifying completeness and purity. In contrast, the selection with the difference of chi squares (v) removes 95\% of the stars at all photometric redshifts, maintaining a solid completeness only lowered by 2\% and even removing extra contaminants.
%
%
We find similar results for the other observing strategies.

From these results, we can conclude that the combination of both the soft (ii) and the soft (v) criteria is the most efficient way of removing stars from the high-redshift galaxy candidates. This is used in the next sections. The remaining stellar contaminants in the $z>4$ galaxy samples decrease from $0.26\pm0.02$ to $0.01\pm0.004$~arcmin$^{-2}$ for CEERS\_1 and from $0.18\pm0.02$ to $0.004\pm0.003$~arcmin$^{-2}$ for HUDF\_1. The differences between CEERS and the HUDF include the input density of stars at the respective sky coordinates, the depth and wavelength coverage of the observations. 
The addition of MIRI imaging improves the photometric redshifts of stars, with detected densities of $0.24\pm0.02$~arcmin$^{-2}$ for CEERS\_2 and $0.14\pm0.02$~arcmin$^{-2}$ for HUDF\_2. These lower values come from the mid-IR colors of stars that are less comparable to galaxy colors than in the near-IR.
It should be mentioned that these selection criteria are specifically constructed to reject stars, nevertheless they are not the only criteria to have this effect. Color-color selections designed to select high-redshift galaxies based on their Lyman break may result in extra stellar rejection, whereas SED-fitting-based selections mainly rely on the types of criteria considered above. The final stellar rejection therefore depends on the entire set of selection criteria.

\subsection{Galaxy selection at $z>5$}

\begin{table*}[t]
\small
\centering
\renewcommand{\arraystretch}{1.1}
\begin{threeparttable}
\caption{Sets of criteria to select high-redshift galaxies. The symbols $\wedge$ and $\vee$ are for logical AND and OR, respectively.}
\begin{tabular*}{\textwidth}{l@{\extracolsep{\fill}}lcl}
\hline\hline
Set & Field && Criteria \\
\hline
Bouwens+2015
&EGS&& $(m_{606}-m_{814}>1.0)$ $\wedge$ $(m_{115}-m_{150}<0.5)$ $\wedge$ $(m_{606}-m_{814}>2.2(m_{115}-m_{150})+1.2)$ \\
&&$\vee$& $(m_{814}-m_{115}>1.1)$ $\wedge$ $(m_{115}-m_{150}<0.6)$ $\wedge$ $(m_{814}-m_{115}>1.1(m_{115}-m_{150})+1.4)$ \\
&&$\vee$& $(m_{115}-m_{150}>1.0)$ $\wedge$ $(m_{150}-m_{200}<0.7)$ $\wedge$ $(m_{115}-m_{150}>0.8(m_{150}-m_{200})+1.0)$ \\
&&$\wedge$& $z_\text{phot}$ in $z_i$ interval \\
&&&\\
&XDF&& $(m_{606}-m_{775}>0.8)$ $\wedge$ $(m_{115}-m_{150}<0.8)$ $\wedge$ $(m_{606}-m_{775}>1.5(m_{115}-m_{150})+1.0)$ \\
&&$\vee$& $(m_{775}-m_{090}>0.6)$ $\wedge$ $(m_{115}-m_{150}<0.8)$ $\wedge$ $(m_{775}-m_{090}>0.8(m_{115}-m_{150})+0.8)$ \\
&&$\vee$& $(m_{090}-m_{115}>0.7)$ $\wedge$ $(m_{115}-m_{150}<1.0)$ $\wedge$ $(m_{090}-m_{115}>0.7(m_{115}-m_{150})+0.9)$ \\
&&$\vee$& $(m_{115}-m_{150}>0.8)$ $\wedge$ $(m_{150}-m_{200}<1.0)$ $\wedge$ $(m_{115}-m_{150}>0.8(m_{150}-m_{200})+0.9)$ \\
&&$\vee$& $(m_{150}-m_{200}>0.8)$ $\wedge$ $(m_{200}-m_{277}<1.0)$ $\wedge$ $(m_{150}-m_{200}>0.8(m_{200}-m_{277})+0.9)$ \\
&&$\wedge$& $z_\text{phot}$ in $z_i$ interval \\
&&&\\
Bowler+2015 
&all&& $z_\text{phot}$ in $z_i$ interval \\
&&$\wedge$& $((z_\text{phot,sec}>z_i-\Delta z)$ $\vee$ $(\chi^2_\text{sec}-\chi^2_\text{gal}>4))$ \\
&&&\\
Finkelstein+2015 
&all&& PDF$z$ integral under primary peak $\ge$ 0.7 \\
&&$\wedge$& PDF$z$ integral in $z_i$ interval $\ge$ 0.25 \\
&&$\wedge$& PDF$z$ integral in $z_i$ interval highest among intervals \\
&&$\wedge$& PDF$z$ integral in $[z_i-1,\infty)$ $\ge$ 0.5 \\
&&$\wedge$& $(z_\text{phot}>z_i-2)$ \\
\hline
\end{tabular*}
\label{tab:selections}
\end{threeparttable}
\end{table*}

\begin{figure*}[t]
  \centering
  \includegraphics[width=0.33\hsize]{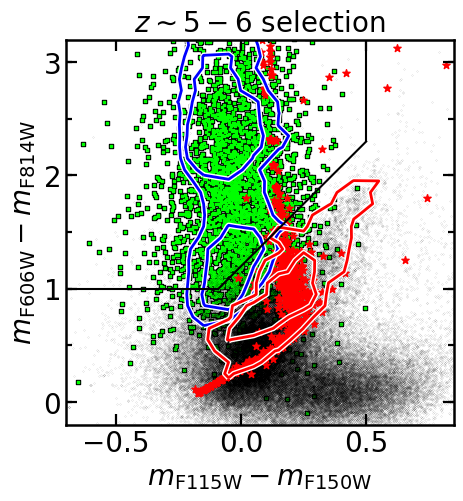}
  \includegraphics[width=0.33\hsize]{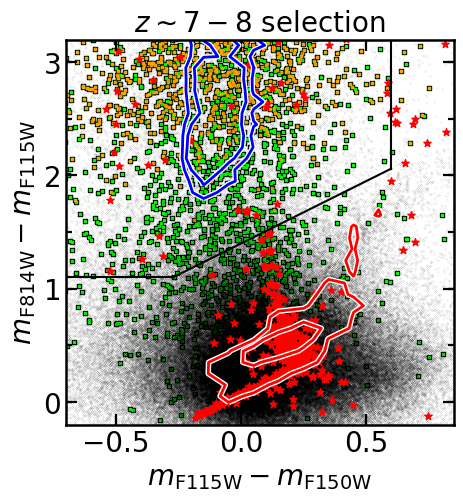}
  \includegraphics[width=0.33\hsize]{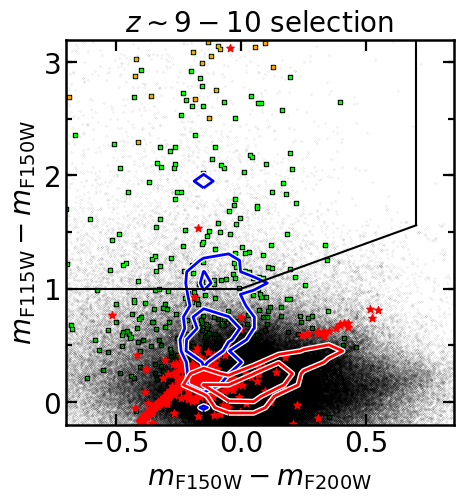}
	\caption{
Color-color selection criteria to preselect galaxies at $z\sim5-6$, $z\sim7-8$, $z\sim9-10$ in the EGS field, with the Bouwens-like criteria in Table~\ref{tab:selections}. The regions enclosed by the solid black line in the top-left corners show the color-color space region in which galaxies are preselected. 
The blue contours enclose 50\% and 80\% of the $z>4.5$, $z>6.5$, $z>8.5$ galaxies input colors (without photometric scatter), and the red contours represent low-redshift quiescent galaxies. Each marker represents the measured colors for a detected source in the CEERS\_1 observing strategy. Only sources detected at $5\sigma$ in the 3, 2, 2 reddest bands, respectively, are indicated. The green and orange squares are $z>4.5$, $z>6.5$, $z>8.5$ galaxies, the orange squares indicating $1\sigma$ upper limits in the bluest band. The red stars are stellar objects, the black dots are low-redshift contaminants. 
	}
	\label{fig_color_color}
\end{figure*}

\begin{figure*}[t]
	\centering
	\includegraphics[width=\hsize]{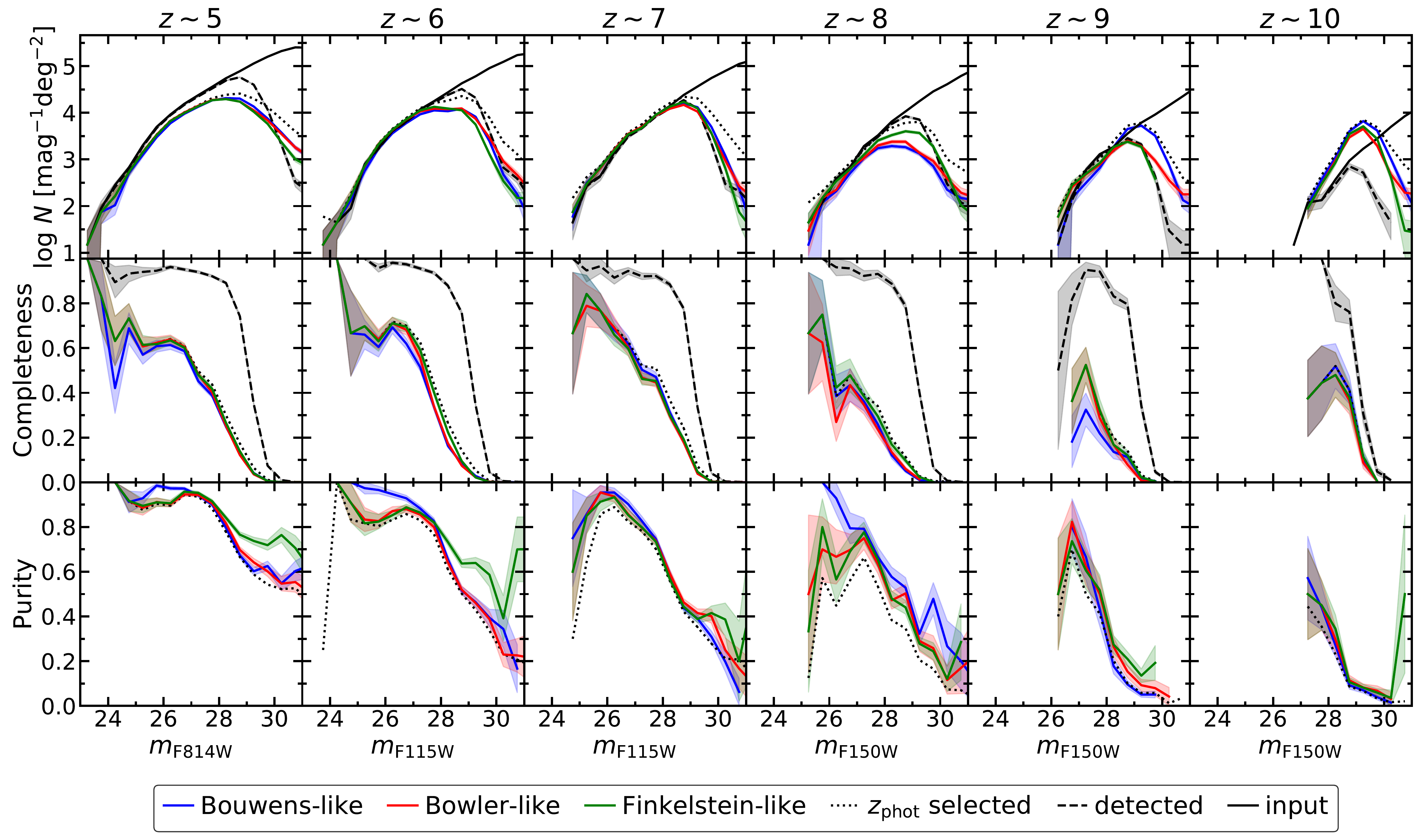}
	\caption{High-redshift galaxy number counts (top panels), completeness (middle panels) and purity (bottom panels) versus apparent magnitude (in the band the nearest to rest-frame UV), in the CEERS\_1 observing strategy. Each column corresponds to one redshift interval. Each colored line represents one set of selection criteria from Table~\ref{tab:selections}. The black solid lines indicate the input number counts. The black dashed lines illustrate detected sources assuming the redshifts are perfectly recovered. The black dotted lines represent observed sources selected with photometric redshifts only. The shaded areas correspond to $1\sigma$ errors. 
	The input and detected counts, and the measured completeness, are expressed in true magnitudes, while the selected counts and purity measurements are in observed magnitudes.
	}
	\label{fig_selection_CP}
\end{figure*}

\begin{figure}[t]
	\centering
	\includegraphics[width=\hsize]{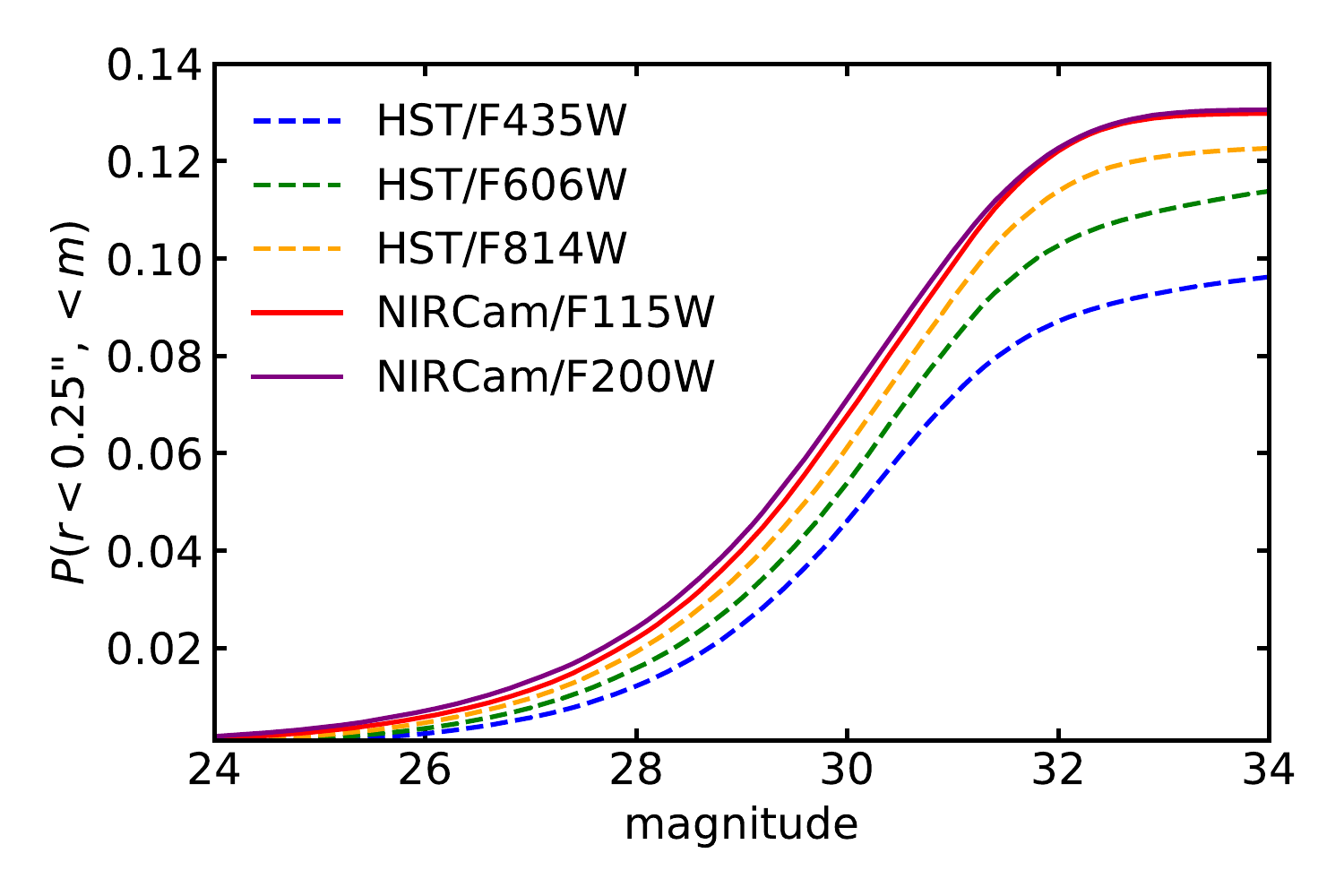}
	\caption{Probability of finding at least one brighter neighbor in the input catalog centered within a 0.25" radius. The line styles and colors represent different bands.}
	\label{fig_probability_brighter_neighbor}
\end{figure}

In this section, we explore multiple procedures to select high-redshift galaxies and estimate the respective impact on galaxy completeness and purity. 
We use an alternative, more permissive definition of purity in this section only. A selected source, assigned to the redshift interval centered at $z_i$, is considered as a contaminant if $z_\text{true}<z_i-1$. Hence only low-redshift sources are considered as contaminants.
We do not treat the specific case of faint Lyman alpha emitters (LAE), typically presenting a strong emission in only one or two bands. The redshift of these galaxies cannot be well constrained without narrow-band imaging or spectroscopy \citep{dunlop_observing_2013}, which are not available here. 
In addition, we do not include criteria based on visual inspection. This technique may be used to discard sources based on shape or colors to consolidate purity, however with real images the resulting galaxy completeness becomes hard to estimate.

We consider three sets of selection criteria summarized in Table~\ref{tab:selections}. These are based on \citet{bouwens_uv_2015}, \citet{bowler_galaxy_2015} and \citet{finkelstein_evolution_2015}, adapted to the present set of photometric bands and generalized to multiple redshift intervals. We do not include magnitude cuts.
The criteria for the EGS field in \citet{bouwens_uv_2015} rely on initial color-color preselections, then on photometric redshifts confirmation. Because of the lack of optical and medium/narrow band imaging, it is not possible to select galaxies with color criteria only. The Lyman break galaxies (LBG) color-color selections are represented in Fig.~\ref{fig_color_color}. The location of the Lyman alpha break relies on the $V_{606}$ and $i_{814}$ \textit{HST} bands at $z<8$. The galaxy colors redward of the break are quantified with NIRCam bands to take advantage of the increased depths compared to the WFC3 bands. Lower-redshift contaminants are expected to be excluded by imposing no detections (SNR<$2\sigma$) blueward of the break. However, this fact is strictly valid at $z>6$ where the IGM transmission is extremely low. The resulting high-redshift samples may be biased toward young UV bright sources and miss a significant fraction of the galaxies \citep{le_fevre_vimos_2015, finkelstein_evolution_2015}, including old or dusty galaxies. Contaminants for high-redshift samples constructed from color-color criteria are usually low-redshift very red dusty galaxies or AGNs, and cool galactic stars. 
In the HUDF field, the deep \textit{HST} optical imaging allows us to develop more redshift-specific color criteria. Nonetheless, we still rely on photometric redshift confirmation for these sources, especially at $z>7$ where the NIRCam broad bands cannot precisely locate the Ly$\alpha$ break. The color criteria for the HUDF field are presented in Appendix~\ref{appendix_galaxy_selection_HUDF}.
Alternatively, \citet{bowler_galaxy_2015} criteria mainly use photometric redshifts and impose additional constraints on the location of the secondary photometric redshift $z_\text{phot,sec}$. 
Similarly, \citet{finkelstein_evolution_2015} criteria make use of the whole posterior information to select objects based on the location and concentration of the PDF$z$ in redshift intervals. In these two approaches, we do not include the criteria on the absolute quality of the galaxy templates fit. Such criteria generally have a marginal impact on the final selection and, in our simulation, may just capture the differences between the input and the fitted templates.

Figure~\ref{fig_selection_CP} illustrates the completeness and purity of the high-redshift galaxy samples in the CEERS\_1 strategy, as a function of apparent observed-frame UV magnitudes $m_\text{UV}$. The colored lines represent the three different selections. The results for photometric redshift only selected sources (dotted lines), and the completeness of the detected sources assuming that redshifts are perfectly recovered (dashed lines), are also shown for comparison. The results for the CEERS\_2 strategy are very similar. 
%

We find that about $5\pm2$\% ($2\pm1$\%) of the bright galaxies at $z=4-6$ ($1-2$) are not detected. This implies that bright nearby objects contaminate the photometry of these sources, for which the source detection or the deblending procedure failed. 
At fainter magnitudes, the drop of completeness is the consequence of both this effect becoming stronger for faint sources and the impact of noise. 
Figure~\ref{fig_probability_brighter_neighbor} illustrates the probability of finding a brighter neighbor in the input source catalog centered within the 0.5"~diameter aperture. In the NIRCam/F200W band, this probability is about 3\% at 28~mag and converges to 13\% at 33~mag. This gives a hint of the impact of blending alone on faint source photometry. Other scenarios are also possible, such as brighter neighbors outside of the aperture dominating the surface brightness of the faint source, therefore undetected or undeblended.

We find that the high-redshift galaxies selected through photometric redshifts only (before applying any other selection) already present significant incompleteness, even at $m_\text{UV}<27$. 
Many sources that are correctly identified as high-redshift galaxies present relatively broad PDF$z$, so the resulting photometric redshifts often reside in the previous or next redshift interval. This is emphasized by the redshift intervals whose widths are fixed and not increasing with redshift. 
The bright high-redshift galaxies with catastrophic photometric redshifts are typically identified as red low-redshift galaxies, however many of them present PDF$z$ with multiple peaks and a correct secondary solution. 
These results reflect the lack of deep optical and/or near-IR medium-band imaging, in the rest-frame UV region of these sources, to better identify the Ly$\alpha$ break, and the lack of blue-band imaging to confirm the break. 
Detected sources with nearby bright extended objects may also present contaminated photometry and colors, even for relatively bright galaxies. 
At very high redshift $z\geq9$, the rarity of the galaxies of interest compared to the significantly more numerous low-redshift contaminants (at $z\sim2$) leads to relatively low purity, in addition to the degeneracy between the Lyman break and the Balmer break at low redshift.

We observe slight differences between the different selection sets with respect to galaxy completeness and purity.
Firstly, the Bouwens-like criteria lead to an improvement in purity, especially at bright magnitudes, with a relatively limited loss of completeness. Photometric redshifts impose most of the constraints, therefore the results are robust against against small changes in the color preselection. Nonetheless, this preselection effectively increases purity, especially at the bright ends. 
Secondly, the Bowler-like selection induces a smaller loss of galaxy completeness, and increases the purity at the faint end.
The criterion on the second peak of the PDF$z$ has a significant impact on both $C$ and $P$, especially at the faint ends.
%
Thirdly, the criteria from Finkelstein lead to the highest galaxy completeness at most magnitudes and redshifts. At the same time, the resulting purity is the highest at the faint ends and at all redshifts, especially at $z<8$. The constraint on the weight of the primary PDF$z$ peak increases the purity and slightly decreases the completeness at the faint end. All the additional criteria increase even more the faint-end purity, however lowering the completeness at bright magnitudes.
With these criteria, we find that the galaxy completeness is higher than $50\%$ for $m_\text{UV}<27.5$ sources at all redshifts, and purity remains above 80 and 60\% at $z\leq7$ and 10 respectively.
%
From this comparison, we conclude that the PDF$z$ criteria of Finkelstein result in the best trade-off between completeness and purity, and we keep these criteria in the next sections. 

Furthermore, completeness and purity may a priori depend on other physical parameters such as galaxy size. 
Completeness is about twice larger for galaxies with effective radius $r_e<0.2$~kpc at $m_\text{UV}>29$ in CEERS and $m_\text{UV}>31$ in the HUDF. These sources are right at the detection limits, where completeness is only a few percent. We observe no significant evidence of purity varying with galaxy size. In the computation of the luminosity function, this variability should be taken into account, although \citet{finkelstein_evolution_2015} notably showed that this has a minor impact.

Figure~\ref{fig_selection_CP_HUDF} illustrates the same analysis in the HUDF\_1 configuration. The completeness after selecting galaxies is much closer to the completeness assuming perfectly recovered redshifts. This is mainly due to the deeper NIRCam imaging and the additional \textit{HST} $B$ band. 
We observe similar features between the three selection sets as for the CEERS configuration.
With the Finkelstein selection, the galaxy completeness remains higher than 50\% at $m_\text{UV}<29$ at all redshifts, and the purity above 80\% at $m_\text{UV}<30$.

\subsection{Number counts predictions}

We quantify the number of detected and selected sources in the high-redshift galaxy samples. Figure~\ref{fig_selection_CP} and \ref{fig_selection_CP_HUDF} show the predicted number counts per magnitude and redshift, for the CEERS\_1 and HUDF\_1 observing strategies. The results are equivalent for the CEERS\_2 and HUDF\_2 configurations, respectively. The selected counts designate the selected objects following the indicated selection and assigned to the corresponding redshift interval. These are computed using observed magnitudes. In contrast, the detected and the input counts are computed using true magnitudes and redshifts. The drop at $m_\text{UV}>31$ at all redshifts comes from the stellar mass lower limit in the input galaxy catalog. The apparent disagreement between the input and the selected counts at $z\sim10$ comes from photometric scatter.

For the CEERS\_1 observing strategy, we expect about 916, 435, 232, 56, 19, 7 true high-redshift galaxies at $m_\text{UV}<29$ that are correctly assigned to the selected samples at $z\sim 5,6,7,8,9,10$ respectively. In comparison, the input number counts are 3039, 1522, 774, 318, 101, 21.
These numbers agree with the predictions from the CEERS program description (20-80 sources at $z=9-13$), though closer to the lower bound. 
One explanation may be source blending and the resulting increase in the photometric redshift outlier rates. Faint sources may even not be detected because of bright nearby objects, especially bright extended galaxies and stars, lowering the detected number counts. In addition, the high-redshift number counts importantly depend on the assumed evolution of the UV LF at $z>8$, so that the rapid evolution assumed here gives lower number counts than with a slower evolution.
For the HUDF\_1 configuration, we expect 205, 135, 65, 20, 6, 2 selected sources at $m_\text{UV}<31$ and $z\sim 5,6,7,8,9,10$ respectively, compared to the 628, 367, 222, 112, 40, 12 input counts.
%
These numbers indicate that the GTO programs in the HUDF are more suitable than CEERS to study very faint galaxies at $z\geq8$, in which case deeper imaging is required. On the other hand, the larger survey area in the EGS field enables more galaxies at $z\sim5-6$ to be detected, including rare intrinsically bright sources.

\subsection{Computing the galaxy luminosity function}

\begin{figure*}[t]
	\centering
	\includegraphics[width=\hsize]{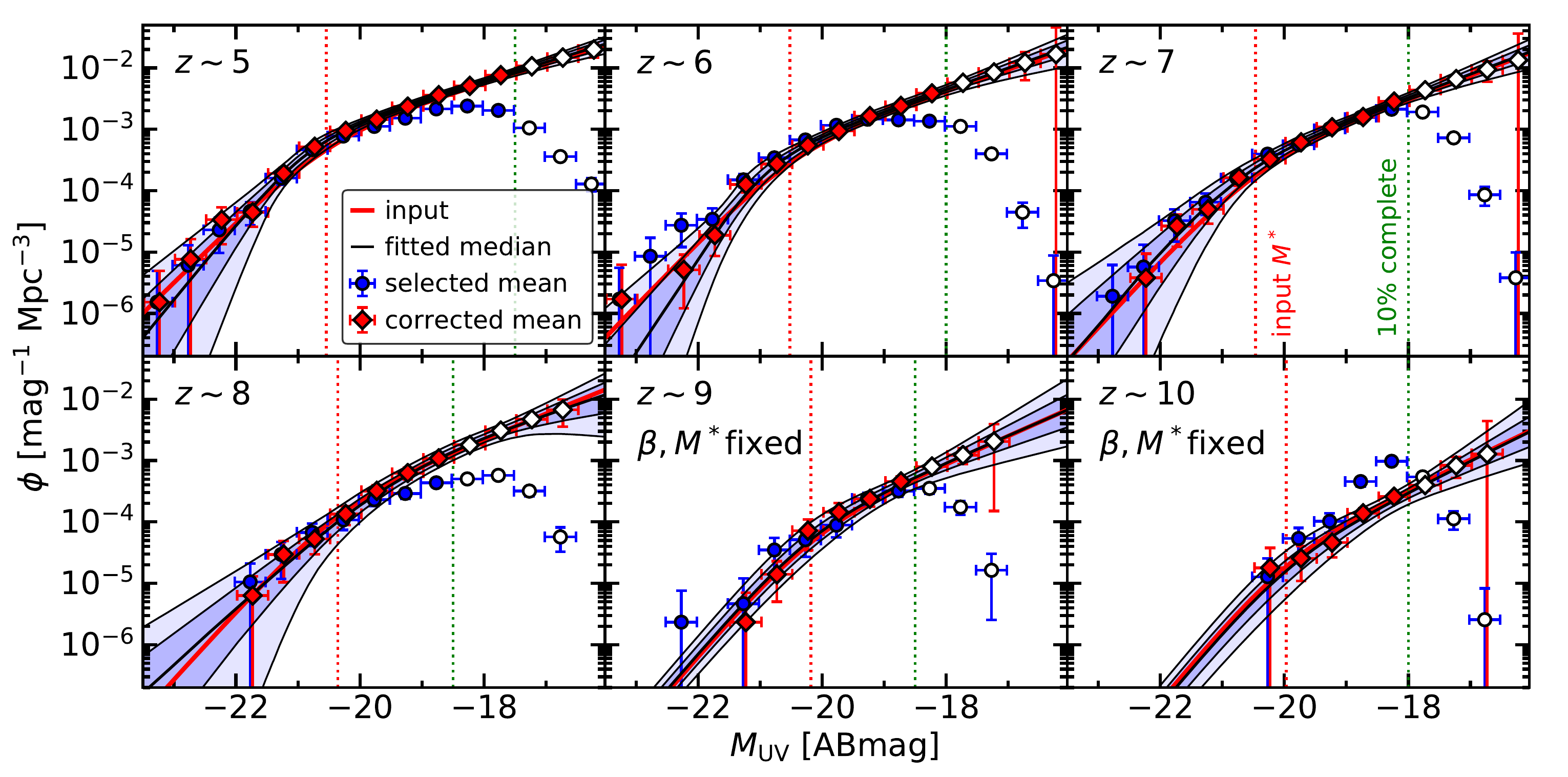}
	\caption{
Galaxy UV luminosity functions for multiple redshift intervals, for the CEERS\_1 observing strategy. The blue dots indicate the estimated mean of the selected counts, and the red diamonds represent the selected counts corrected for incompleteness and impurity. The open symbols are the points where the completeness is below 10\%. The error bars are Poisson errors for the former and the quadratic sum of Poisson and scaling errors for the latter. Multiple Poisson random vectors are sampled from the blue dots, scaled to correct for incompleteness and impurity and fitted with a double power-law function, with the indicated fixed parameters. The black lines show the model with the median fitted parameters, after marginalizing over all the sampled counts. The colored areas indicate 1$\sigma$ and 2$\sigma$ credibility errors. The red lines represent the input luminosity functions (fitted with a DPL). The red dotted lines show $M^*$ from the input LF, and the green dotted lines indicate the 10\% completeness limit.
	}
	\label{fig_LF}
\end{figure*}

In this section, we discuss the computation of the galaxy UV luminosity function (LF) from the selected galaxy number counts and the measured completeness and purity. 
The luminosity function is the comoving volume number density as a function of the intrinsic luminosity. 
The observed number density may suffer from incompleteness and impurities, therefore the observed LF needs to be corrected to recover the intrinsic LF using magnitude-dependent scaling factors.

The input galaxy UVLF in JAGUAR is constructed from the convolution of the stellar mass function and the $M_\text{UV}(M_*)$ relation.
Because of the stellar mass lower limit $\log(M_*/M_\odot)>6$,
the LF decreases at the faint end with a maximum situated between $-16<M_\text{UV}<-15$ at $4<z<10$. The position of this turn-over is still debated in the literature \citep[e.g.,][]{livermore_directly_2017,bouwens_z_2017}, therefore we restrict ourselves to $M_\text{UV}<-16$ where the faint end remains almost linear.
We fit this input UVLF at $M_\text{UV}>-22$ with a double-power-law model (DPL), parametrized as \citep{bowler_galaxy_2015}:
\begin{equation}
\phi(M) = \dfrac{\phi^*}{10^{0.4(\alpha+1)(M-M^*)}+10^{0.4(\beta+1)(M-M^*)}},
\end{equation}
where $\phi^*$ and $M^*$ are the characteristic density and magnitude, $\alpha$ and $\beta$ denote the faint and bright-end slopes.
The difference between the input UVLF and the fitted model at $z\leq10$ is at most 10\% between $-22<M_\text{UV}<-17$.

We make forecasts for the recovery of the UVLF with the following approach. 
We take the selected galaxy $M_\text{UV}$ number counts from our simulations, multiply them by the ratio of the survey area to the simulated area, and sample Poisson random vectors taking these values as the mean. The sampled counts are then corrected for incompleteness and impurities through the scaling correction factors, estimated from the number of input sources (function of true magnitudes) divided by the number of selected objects (function of observed magnitudes) from our simulations. This scaling therefore includes photometric scatter and $M_\text{UV}$ recovery. 
We recall that absolute magnitudes are not corrected for dust attenuation. 
We use the classic estimator of the LF \citep{felten_schmidts_1976}, consisting of the absolute UV magnitude number counts divided by the comoving volume in the whole redshift interval. 
The LF uncertainties are the quadratic sum of the Poisson errors, cosmic variance errors \citep{trenti_cosmic_2008} and scaling correction uncertainties. 
By construction, the corrected LF values equal the input LF ones, however the uncertainties are broadened depending on the selected sample sizes. 
We fit each scaled Poisson random vector at $M_\text{UV}<-16$ with a DPL model, using flat priors and a Markov Chain Monte Carlo (MCMC) method to sample the posterior probability distribution. We finally marginalize over the Poisson samplings to determine the median parameters and errors.
Both the statistical and the systematic errors on the parameters are included, though in reality one would have one Poisson sampling and only determine the statistical errors.

The recovered LFs are presented in Fig.~\ref{fig_LF} and Table~\ref{tab:CEERS-counts} for the CEERS\_1 configuration covering about 100~arcmin$^2$. We do not present the results for the HUDF strategy, because the 4.7~arcmin$^2$ survey area cannot impose much statistical constraint on the LF, despite the increased depths. 
%
The differences between the selected and the corrected counts are significant, especially beyond $M_\text{UV}>-18$ where the galaxy samples become highly incomplete. At $M_\text{UV}<-18$, the scaling corrections are still $\sim1-2$ at all redshifts. 
Poisson uncertainties dominate the LF error budget at the bright end where the number counts are low, while cosmic variance errors reaches up to 70\% of the total variance at fainter magnitudes. Scaling corrections contribute to about $10$\% of the total variance at almost all magnitudes and redshifts. 
As with real images, the corrections remain strongly dependent on the modelling assumptions, including galaxy morphology, star-formation histories and dust attenuation.
These results reflect that accurate simulations are required to correctly recover the galaxy counts, that can be severely affected by incompleteness and contamination. 
%
The number counts brightward of $M^*$ decrease with increasing redshift, leading to a lack of constraints on the bright end. 
For this reason, we fix the DPL parameters $\beta$ and $M^*$, at $z\geq9$, to the input values when performing the fit. The obtained parameters are presented in Table~\ref{tab:LF-params}. 
The faint-end slopes are effectively constrained with an absolute error of $\sim0.1$ at $z\leq7$ and $\sim0.25$ at $z\geq8$.

Within the CEERS area of 100~arcmin$^2$, the input galaxy UVLF predicts about 71, 36, 19, 12, 3.3 and 1.3 input galaxies with $M_\text{UV}<M^*$ at $z\sim5,6,7,8,9,10$ respectively. 
These numbers indicate that the bright end of the UVLF cannot be constrained at $z\ge7$, even assuming that all these galaxies are identified. Nonetheless, the NIRCam GTO program in the GOODS fields covering 200~arcmin$^2$, particularly the "Medium" survey, will bring additional constraints on the bright end of the UVLF up to $z\leq8$.
In spite of the depths of these programs, the main limitation remains the small \textit{JWST} field of view. As an alternative, the Euclid\footnote{\url{http://www.euclid-ec.org}} deep fields will include optical and near-IR imaging extended over tens of square degrees \citep{laureijs_euclid_2011}. These surveys, with the optical (e.g., Subaru Hyper Suprime-Cam) and mid-infrared (e.g., Spitzer Legacy Survey) counterparts, will reach the required depth to identify high-redshift galaxies, despite a lower resolution than \textit{JWST}. The Euclid deep fields will probe the bright end of the luminosity function up to $z\sim7$ or more, which will provide constraints complementary to the deep \textit{JWST} surveys.



In addition, we predict the recovery of the cosmic SFR density $\rho_\text{SFR}$. We integrate the UVLF to $M_\text{UV}=-16$. The UV luminosity densities are converted into SFR densities using $\kappa_\text{UV}=1.15\times10^{-28}$~$M_\odot$yr$^{-1}$(erg/s/Hz)$^{-1}$, where a 0.1-100~$M_\odot$ Salpeter initial mass function and a constant SFR are assumed \citep{madau_cosmic_2014}. The results, uncorrected for dust attenuation, are reported in Table~\ref{tab:LF-params}.
The SFR densities are correctly recovered and the expected errors remain below 0.1~dex, as long as the faint-end slope is well constrained. However, the errors are underestimated because of the fixed LF parameters at $z\ge9$, and the scaling corrections recovering the input number counts. In addition, we do not apply any magnitude cuts, which would significantly lower the number of faint selected sources.
In the ideal case where all the detected sources have perfectly recovered redshifts and absolute magnitudes, the errors on $\alpha$ and $\rho_\text{SFR}$ are lowered by about 20\% at $z<8$. The cases at $z>8$ are more sensitive to the determination of $\alpha$ from small number counts at the very faint end.
Using all the input sources over the survey area at $\log(M_*/M_\odot)>6$, we estimate that about 50\% of the total errors arise from the limited area. This argues again in favor of surveys including larger cosmological volumes.

\begin{table}
\footnotesize
\setlength{\tabcolsep}{2pt}
\centering
\begin{threeparttable}
\caption{Parametric fitting of the recovered UVLF}
\begin{tabular}{cccccc}
\hline\hline
$z$ & $\phi^*$ & $M^*$ & $\alpha$ & $\beta$ & $\log\rho_\text{SFR}$ \\
& [$10^{-3}$Mpc$^{-3}$] & [mag] & & & [$M_\odot$yr$^{-1}$Mpc$^{-3}$] \\\hline
& \multicolumn{5}{c}{input} \\
5 & 0.92 & $-20.54$ & $-1.78$ & $-3.50$ & $-1.64$ \\
6 & 0.55 & $-20.52$ & $-1.87$ & $-3.63$ & $-1.80$ \\
7 & 0.35 & $-20.46$ & $-1.96$ & $-3.73$ & $-1.96$ \\
8 & 0.24 & $-20.36$ & $-2.03$ & $-3.79$ & $-2.11$ \\
9 & 0.09 & $-20.18$ & $-2.13$ & $-3.95$ & $-2.53$ \\
10 & 0.04 & $-19.97$ & $-2.22$ & $-4.07$ & $-2.96$ \\
& \multicolumn{5}{c}{recovered} \\
5 & $0.80_{-0.27}^{+0.54}$ & $-20.84_{-0.27}^{+0.39}$ & $-1.77_{-0.08}^{+0.10}$ & $-4.10_{-1.21}^{+0.71}$ & $-1.57_{-0.03}^{+0.03}$ \\[3pt]
6 & $0.42_{-0.17}^{+0.44}$ & $-20.85_{-0.35}^{+0.48}$ & $-1.89_{-0.11}^{+0.15}$ & $-4.79_{-1.56}^{+1.00}$ & $-1.75_{-0.03}^{+0.03}$ \\[3pt]
7 & $0.32_{-0.17}^{+0.39}$ & $-20.67_{-0.53}^{+0.55}$ & $-1.94_{-0.11}^{+0.15}$ & $-3.93_{-1.18}^{+0.73}$ & $-1.89_{-0.04}^{+0.05}$ \\[3pt]
8 & $0.96_{-0.76}^{+2.45}$ & $-19.39_{-1.05}^{+1.04}$ & $-1.84_{-0.28}^{+0.49}$ & $-3.27_{-1.13}^{+0.43}$ & $-2.13_{-0.07}^{+0.07}$ \\[3pt]
9 & $0.10_{-0.03}^{+0.04}$ & $-20.18$ & $-2.09_{-0.22}^{+0.24}$ & $-3.95$ & $-2.51_{-0.10}^{+0.10}$ \\[3pt]
10 & $0.03_{-0.01}^{+0.02}$ & $-19.97$ & $-2.25_{-0.27}^{+0.25}$ & $-4.07$ & $-3.00_{-0.09}^{+0.10}$ \\[3pt]
\hline
\end{tabular}
\label{tab:LF-params}
\end{threeparttable}
\end{table}

\section{Summary and conclusion}
\label{section6}

In this paper, we forecast the performance of accepted \textit{JWST} deep imaging surveys regarding the detection and analysis of high-redshift galaxies. In particular, we estimate the galaxy physical parameters, optimize the candidate selection with respect to galaxy completeness, purity and the total number of sources, then compute the UV luminosity function and the cosmic star-formation rate density. We treat two \textit{JWST} imaging programs, including CEERS in the EGS field, and HUDF GTO, and simulate the ancillary \textit{HST} data for these fields.
We construct complete mock samples of galaxies, local stars and brown dwarfs, representative of the current understanding of these populations using the latest observed luminosity and mass functions extrapolated to low masses, and high redshifts. The photometry of these sources is simulated through astronomical image generation, following the current knowledge of the \textit{JWST} instruments. We extract the sources with SExtractor and estimate the source physical properties using SED-fitting. 


Our main results can be summarized as follows:
\begin{itemize}

  \item We find that the photometric redshifts estimated in the CEERS configuration are mainly limited by the lack of blue-band data. The additional MIRI bands marginally improve the photometric redshifts at faint magnitudes and at high redshift, where MIRI covers the rest-frame optical. Source blending contributes to up to 20\% of the photometric redshift outliers in CEERS, and 40\% in the HUDF.

  \item Stellar masses are recovered within 0.2~dex at $z\leq5$ and 0.25~dex at $z>5$, and are systematically overestimated by 0.1~dex at high redshift. Star-formation rates are scattered over 0.3~dex and the most star-forming galaxies have a systematic bias of 0.1 to 0.2~dex. Numerous catastrophic SFR estimates arise from photometric redshift outliers.
  
  \item Galactic brown dwarfs contaminating the $z\geq5$ galaxy samples can be effectively discarded, reaching a residual density of $<0.01$~arcmin$^{-2}$. 
  The impact on galaxy completeness remains minimal, although dependent on the assumed galaxy morphology.
  
  \item We find that the $5<z<10$ galaxy selection based on the redshift posterior probability distribution from SED-fitting gives the best compromise between completeness and purity. In the CEERS configuration, galaxy completeness remains above $50\%$ at $m_\text{UV}<27.5$ and purity is higher than 80 and 60\% at $z\leq7$ and 10 respectively. In the HUDF strategy, the galaxy samples are more than $50\%$ complete at $m_\text{UV}<29$ and $80\%$ pure at $m_\text{UV}<30$ at all redshifts.
  
  \item We provide scaling correction factors for the selected galaxy number counts to recover the intrinsic number counts in the CEERS configuration. The values typically range from 1 to 2 at $M_\text{UV}<-18$, but increase a lot at fainter magnitudes. This scaling is sensitive to the source modeling used as input, the source extraction and template fitting procedure, as well as the choice of ancillary data. Thus, the provided factors are strictly valid when using the same procedure presented here. However, our results show how crucial these types of calculations are to correctly recovering the luminosity function.
  
  \item The faint-end slope of the galaxy UV luminosity function in CEERS can be recovered with an error of $\pm0.1$ at $z=5$ and $\pm0.25$ at $z=10$, despite the significant dependence on the correction factors. We estimate that at least 300~arcmin$^2$ would be necessary to constrain the bright end up to $z=8$. 
  
\end{itemize}

We remind the reader that our forecasts are based on future \textit{JWST} and existing \textit{HST} imaging data, meaning that we neglect ancillary spectroscopy and ground-based imaging that may improve the results. In addition, the UVCANDELS program will enlarge the wavelength coverage in the EGS field, which may significantly improve the estimated photometric redshifts and the purity of the high-redshift galaxy samples.

In the future, we plan to include more realistic galaxy morphologies and use our simulations to fully exploit data from \textit{JWST} imaging surveys. In addition, we plan to extend our simulations to the Euclid deep fields.


\begin{acknowledgements}
	CCW acknowledges support from the National Science Foundation Astronomy and Astrophysics Fellow-ship grant AST-1701546. 
	ECL acknowledges support from the ERC Advanced Grant 695671 'QUENCH'. 
	LC acknowledges support from the Spanish Ministry for Science and  Innovation under grants ESP2017-83197 and MDM-2017-0737 "Unidad de Excelencia María de Maeztu - Centro de  Astrobiología (CSIC-INTA)". 
	JPP acknowledges the UK Science and Technology Facilities Council and the UK Space Agency for their support of the UK's JWST MIRI development activities. 
	KIC acknowledges funding from the European Research Council through the award of the Consolidator Grant ID 681627-BUILDUP.
\end{acknowledgements}

\bibliography{main_arxiv}
\bibliographystyle{mnras} 

\onecolumn
\appendix
\addcontentsline{toc}{section}{Appendix}

\section{Tables}

\begin{table*}[h]
\footnotesize
\centering
\caption{Galaxy absolute magnitude number counts for luminosity function computation in CEERS\_1 observing strategy. Per columns, the expectation value of the selected number counts, completeness, purity and scaling correction factor.}
\begin{minipage}[t]{.49\linewidth}
\begin{tabular}[t]{c|cccc}
\hline\hline
$M_\text{UV}$ & $\mathbb{E}[N]$ & $C$ & $P$ & $S$ \\\hline
& \multicolumn{4}{c}{$z\sim5$} \\
-22.75 & $0.8\pm0.2$ & $0.8\pm0.2$ & $1\pm0$ & $1.2\pm0.3$ \\
-22.25 & $3.0\pm0.3$ & $0.7\pm0.1$ & $0.93\pm0.06$ & $1.5\pm0.2$ \\
-21.75 & $6.0\pm0.5$ & $0.83\pm0.07$ & $0.83\pm0.07$ & $1.0\pm0.1$ \\
-21.25 & $21.0\pm0.9$ & $0.76\pm0.04$ & $0.83\pm0.04$ & $1.19\pm0.08$ \\
-20.75 & $61\pm2$ & $0.67\pm0.03$ & $0.75\pm0.02$ & $1.11\pm0.05$ \\
-20.25 & $101\pm2$ & $0.64\pm0.02$ & $0.84\pm0.02$ & $1.22\pm0.04$ \\
-19.75 & $146\pm2$ & $0.62\pm0.02$ & $0.85\pm0.01$ & $1.29\pm0.03$ \\
-19.25 & $198\pm3$ & $0.52\pm0.01$ & $0.79\pm0.01$ & $1.54\pm0.04$ \\
-18.75 & $281\pm3$ & $0.44\pm0.01$ & $0.73\pm0.01$ & $1.68\pm0.04$ \\
-18.25 & $313\pm4$ & $0.283\pm0.008$ & $0.62\pm0.01$ & $2.13\pm0.05$ \\
-17.75 & $266\pm3$ & $0.149\pm0.005$ & $0.48\pm0.01$ & $3.8\pm0.1$ \\
-17.25 & $137\pm2$ & $0.044\pm0.002$ & $0.39\pm0.02$ & $10.2\pm0.4$ \\
-16.75 & $47\pm1$ & $<0.01$ & $0.42\pm0.03$ & $41\pm3$ \\
-16.25 & $16.8\pm0.8$ & $<0.01$ & $0.35\pm0.05$ & $155\pm17$ \\
& \multicolumn{4}{c}{$z\sim6$} \\
-22.25 & $3.2\pm0.4$ & $1\pm0$ & $0.2\pm0.1$ & $0.2\pm0.1$ \\
-21.75 & $4.0\pm0.4$ & $0.6\pm0.1$ & $0.6\pm0.1$ & $0.6\pm0.1$ \\
-21.25 & $17.6\pm0.8$ & $0.70\pm0.05$ & $0.67\pm0.05$ & $0.84\pm0.07$ \\
-20.75 & $40\pm1$ & $0.66\pm0.04$ & $0.63\pm0.03$ & $0.80\pm0.05$ \\
-20.25 & $78\pm2$ & $0.68\pm0.03$ & $0.68\pm0.02$ & $0.82\pm0.03$ \\
-19.75 & $136\pm2$ & $0.69\pm0.02$ & $0.62\pm0.02$ & $0.81\pm0.03$ \\
-19.25 & $171\pm3$ & $0.53\pm0.02$ & $0.54\pm0.02$ & $1.12\pm0.04$ \\
-18.75 & $166\pm3$ & $0.32\pm0.01$ & $0.47\pm0.02$ & $1.69\pm0.06$ \\
-18.25 & $159\pm3$ & $0.169\pm0.008$ & $0.40\pm0.02$ & $2.8\pm0.1$ \\
-17.75 & $130\pm2$ & $0.062\pm0.004$ & $0.33\pm0.02$ & $5.2\pm0.2$ \\
-17.25 & $47\pm1$ & $<0.01$ & $0.30\pm0.03$ & $21\pm1$ \\
-16.75 & $5.2\pm0.5$ & $<0.01$ & $0.35\pm0.09$ & $275\pm54$ \\
-16.25 & $0.4\pm0.1$ & $<0.01$ & $0.5\pm0.4$ & $4913\pm3474$ \\
& \multicolumn{4}{c}{$z\sim7$} \\
-22.25 & $0.6\pm0.2$ & $1\pm0$ & $0.3\pm0.3$ & $0.7\pm0.5$ \\
-21.75 & $3.4\pm0.4$ & $0.86\pm0.09$ & $0.88\pm0.08$ & $0.82\pm0.05$ \\
-21.25 & $6.8\pm0.5$ & $0.85\pm0.07$ & $0.56\pm0.09$ & $0.8\pm0.1$ \\
-20.75 & $16.8\pm0.8$ & $0.65\pm0.05$ & $0.65\pm0.05$ & $1.02\pm0.09$ \\
-20.25 & $42\pm1$ & $0.60\pm0.04$ & $0.58\pm0.03$ & $0.82\pm0.05$ \\
-19.75 & $58\pm2$ & $0.50\pm0.03$ & $0.51\pm0.03$ & $1.10\pm0.06$ \\
-19.25 & $108\pm2$ & $0.44\pm0.02$ & $0.45\pm0.02$ & $1.05\pm0.05$ \\
-18.75 & $164\pm3$ & $0.34\pm0.02$ & $0.32\pm0.02$ & $1.01\pm0.04$ \\
-18.25 & $223\pm3$ & $0.22\pm0.01$ & $0.22\pm0.01$ & $1.33\pm0.05$ \\
-17.75 & $201\pm3$ & $0.063\pm0.005$ & $0.19\pm0.01$ & $2.27\pm0.08$ \\
-17.25 & $76\pm2$ & $<0.01$ & $0.21\pm0.02$ & $9.0\pm0.5$ \\
-16.75 & $9.2\pm0.6$ & $<0.01$ & $0.15\pm0.05$ & $108\pm16$ \\
-16.25 & $0.4\pm0.1$ & $<0.01$ & $<0.01$ & $3512\pm2483$ \\
\hline
\end{tabular}
\end{minipage}
\begin{minipage}[t]{.49\linewidth}
\begin{tabular}[t]{c|cccc}
\hline\hline
$M_\text{UV}$ & $\mathbb{E}[N]$ & $C$ & $P$ & $S$ \\\hline
& \multicolumn{4}{c}{$z\sim8$} \\
-21.75 & $1.0\pm0.2$ & $0.7\pm0.3$ & $0.6\pm0.2$ & $0.6\pm0.2$ \\
-21.25 & $2.8\pm0.3$ & $0.6\pm0.1$ & $0.6\pm0.1$ & $1.0\pm0.2$ \\
-20.75 & $6.4\pm0.5$ & $0.5\pm0.1$ & $0.44\pm0.09$ & $0.8\pm0.1$ \\
-20.25 & $10.4\pm0.6$ & $0.45\pm0.06$ & $0.58\pm0.07$ & $1.2\pm0.2$ \\
-19.75 & $21.8\pm0.9$ & $0.39\pm0.04$ & $0.53\pm0.05$ & $1.4\pm0.1$ \\
-19.25 & $28\pm1$ & $0.23\pm0.02$ & $0.44\pm0.04$ & $2.1\pm0.2$ \\
-18.75 & $41\pm1$ & $0.16\pm0.02$ & $0.26\pm0.03$ & $2.5\pm0.2$ \\
-18.25 & $48\pm1$ & $0.060\pm0.008$ & $0.21\pm0.03$ & $3.6\pm0.2$ \\
-17.75 & $54\pm1$ & $0.014\pm0.003$ & $0.14\pm0.02$ & $5.4\pm0.3$ \\
-17.25 & $30\pm1$ & $<0.01$ & $0.09\pm0.02$ & $15\pm1$ \\
-16.75 & $5.4\pm0.5$ & $<0.01$ & $0.07\pm0.05$ & $118\pm23$ \\
-16.25 & $<0.01$ & $<0.01$ & $<0.01$ & $<0.01$ \\
& \multicolumn{4}{c}{$z\sim9$} \\
-21.25 & $0.4\pm0.1$ & $<0.01$ & $<0.01$ & $0.5\pm0.6$ \\
-20.75 & $3.0\pm0.3$ & $0.5\pm0.2$ & $0.5\pm0.1$ & $0.4\pm0.1$ \\
-20.25 & $4.4\pm0.4$ & $0.52\pm0.09$ & $0.6\pm0.1$ & $1.4\pm0.3$ \\
-19.75 & $7.6\pm0.6$ & $0.35\pm0.06$ & $0.37\pm0.08$ & $1.6\pm0.3$ \\
-19.25 & $20.8\pm0.9$ & $0.27\pm0.04$ & $0.31\pm0.05$ & $1.0\pm0.1$ \\
-18.75 & $27\pm1$ & $0.12\pm0.02$ & $0.16\pm0.03$ & $1.4\pm0.1$ \\
-18.25 & $30\pm1$ & $0.05\pm0.01$ & $0.12\pm0.03$ & $2.3\pm0.2$ \\
-17.75 & $15.0\pm0.8$ & $<0.01$ & $0.11\pm0.04$ & $7.0\pm0.9$ \\
-17.25 & $1.4\pm0.2$ & $<0.01$ & $<0.01$ & $126\pm48$ \\
-16.75 & $<0.01$ & $<0.01$ & $<0.01$ & $<0.01$ \\
-16.25 & $<0.01$ & $<0.01$ & $<0.01$ & $<0.01$ \\
& \multicolumn{4}{c}{$z\sim10$} \\
-20.25 & $1.0\pm0.2$ & $0.3\pm0.2$ & $0.4\pm0.2$ & $1.4\pm0.7$ \\
-19.75 & $4.2\pm0.4$ & $0.4\pm0.2$ & $0.3\pm0.1$ & $0.5\pm0.1$ \\
-19.25 & $8.2\pm0.6$ & $0.6\pm0.1$ & $0.22\pm0.06$ & $0.4\pm0.1$ \\
-18.75 & $36\pm1$ & $0.43\pm0.07$ & $0.08\pm0.02$ & $0.30\pm0.04$ \\
-18.25 & $76\pm2$ & $0.16\pm0.04$ & $0.04\pm0.01$ & $0.26\pm0.03$ \\
-17.75 & $42\pm1$ & $<0.01$ & $0.03\pm0.01$ & $0.74\pm0.08$ \\
-17.25 & $8.8\pm0.6$ & $<0.01$ & $0.05\pm0.03$ & $7\pm1$ \\
-16.75 & $0.20\pm0.09$ & $<0.01$ & $<0.01$ & $501\pm501$ \\
-16.25 & $<0.01$ & $<0.01$ & $<0.01$ & $<0.01$ \\
\hline
\end{tabular}
\end{minipage}
\label{tab:CEERS-counts}
\end{table*}

\section{Galaxy selection at $z>5$ in the HUDF}
\label{appendix_galaxy_selection_HUDF}

\begin{figure*}[h]
  \centering
  \includegraphics[width=0.3\hsize]{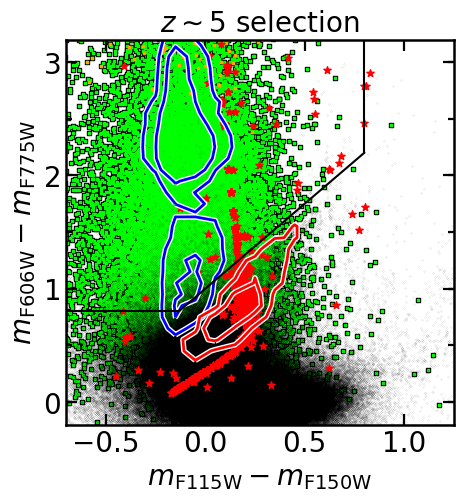}
  \includegraphics[width=0.3\hsize]{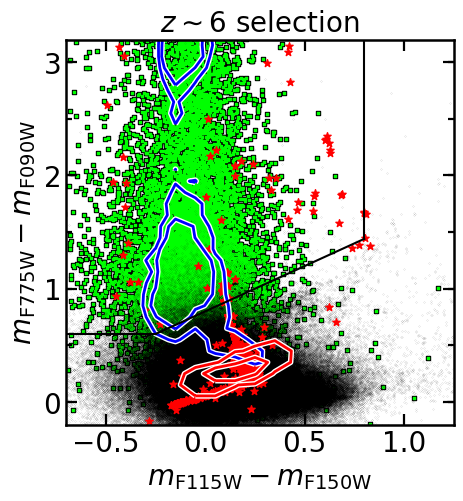}
  \includegraphics[width=0.3\hsize]{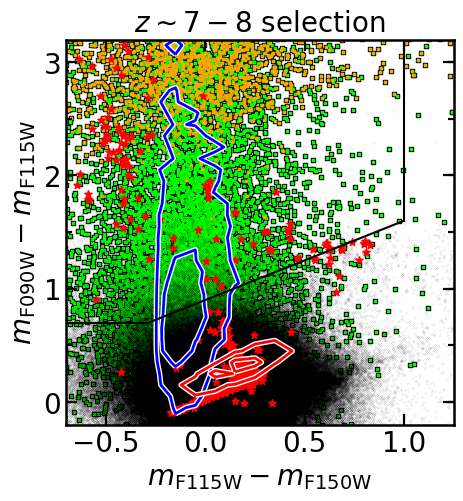}
  \includegraphics[width=0.3\hsize]{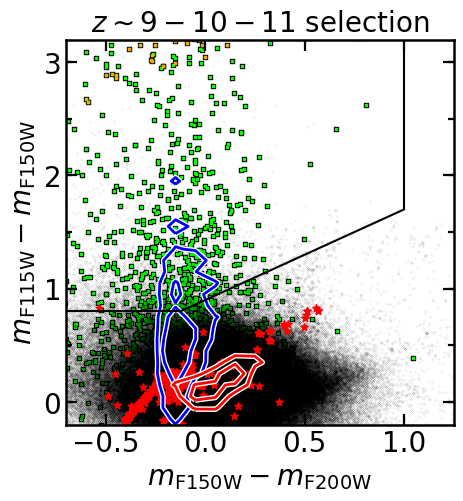}
  \includegraphics[width=0.3\hsize]{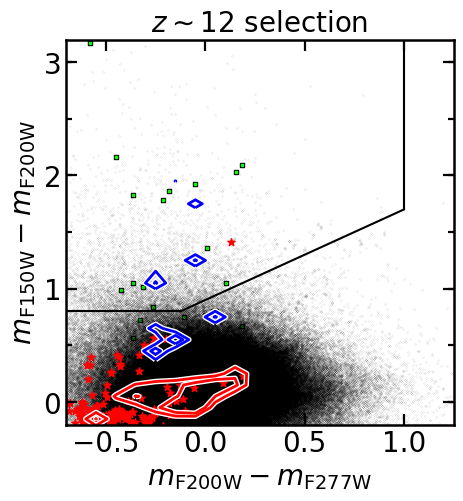}
	\caption{Same as Fig.~\ref{fig_color_color} in the HUDF, for the HUDF\_1 observing strategy.}
	\label{fig_color_color_HUDF}
  \vspace{1em}
	\centering
	\includegraphics[width=\hsize]{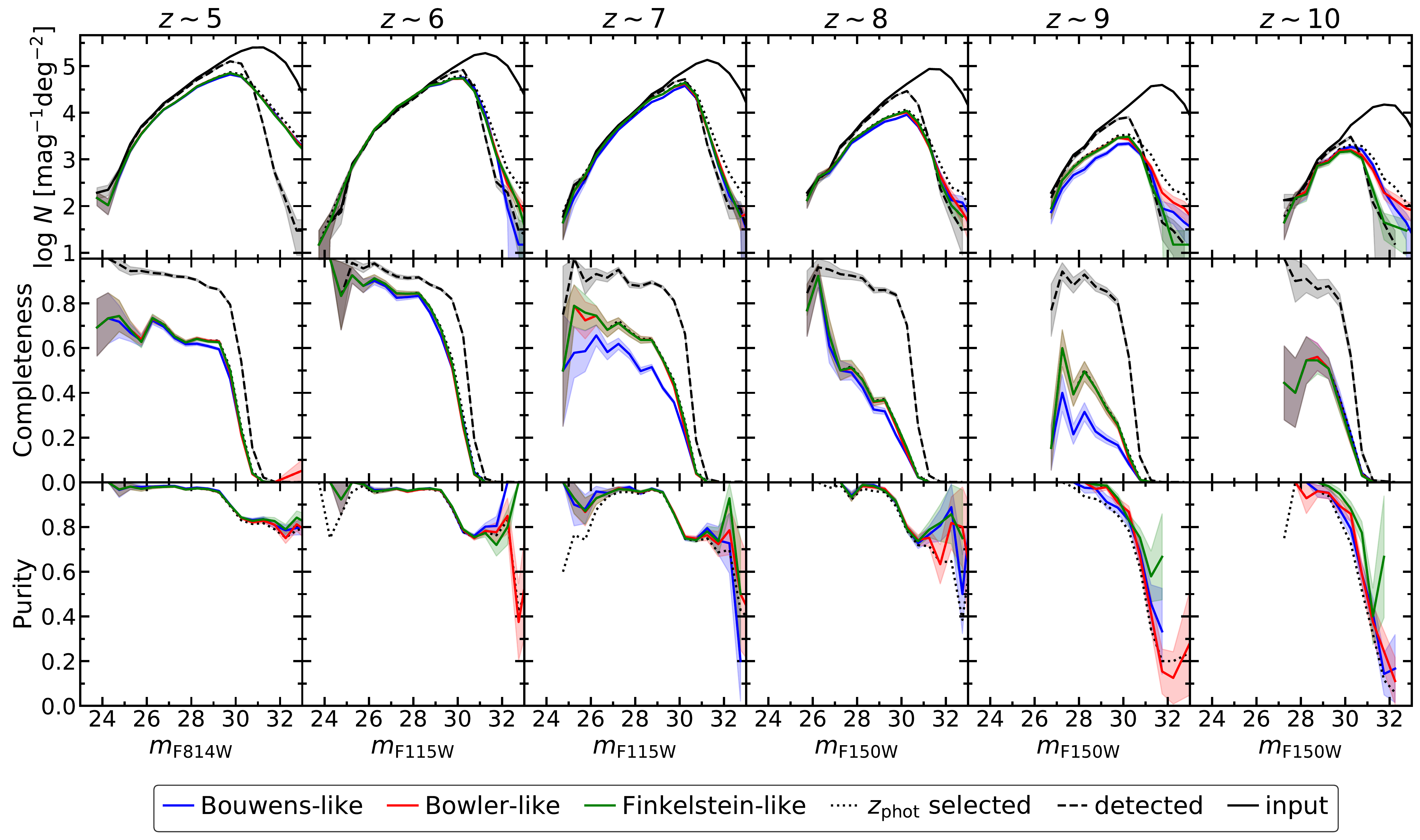}
	\caption{Same as Fig.~\ref{fig_selection_CP} for the HUDF\_1 observing strategy.}
	\label{fig_selection_CP_HUDF}
\end{figure*}

To select our high-redshift galaxies in the HUDF in the Bouwens-like set of criteria, we first preselect sources at $z\sim5$ to 12 using the selection criteria in Table~\ref{tab:selections}. These color criteria are represented in Fig.~\ref{fig_color_color_HUDF}. The high-redshift candidates are then confirmed with photometric redshifts. Figure~\ref{fig_selection_CP_HUDF} indicates the high-redshift galaxy completeness and purity for the Bouwens-, Bowler- and Finkelstein-like criteria. 

\end{document}